\documentclass[fleqn,usenatbib]{mnras}

\usepackage{newtxtext,newtxmath}
\usepackage[T1]{fontenc}

\DeclareRobustCommand{\VAN}[3]{#2}
\let\VANthebibliography\thebibliography
\def\thebibliography{\DeclareRobustCommand{\VAN}[3]{##3}\VANthebibliography}

\usepackage{graphicx}	% Including figure files
\usepackage{amsmath}	% Advanced maths commands
\usepackage{xspace}
\newcommand{\fOtwo}{\textit{f}O$_2$\xspace}

\title[Lava ocean Worlds in high res.]{Detectability of oxygen fugacity regimes in the magma ocean world 55 Cancri e at high spectral resolution}

\author[S. Dash et al.]{Spandan Dash,$^{1,2}$\thanks{E-mail: dashspandan@gmail.com}
Matteo Brogi,$^{3}$
Fabian Lukas Seidler,$^{4}$
Paolo A. Sossi,$^{4}$
Siddharth Gandhi,$^{1,2}$
\newauthor
Vatsal Panwar,$^{1,2}$
Marina Lafarga,$^{1,2}$
and Peter J.\ Wheatley$^{1,2}$
\\
$^{1}$Department of Physics, University of Warwick, Coventry CV4 7AL, United Kingdom\\
$^{2}$Centre for Exoplanets and Habitability, University of Warwick, Coventry, CV4 7AL, United Kingdom\\
$^{3}$Department of Physics, University of Turin, Via Pietro Giuria 1, I-10125, Turin, Italy\\
$^{4}$Institute of Geochemistry and Petrology, ETH Zurich, Clausiusstrasse 25, 8092 Zurich, Switzerland
}

\date{Accepted XXX. Received YYY; in original form ZZZ}

\pubyear{2024}

\begin{document}
\label{firstpage}
\pagerange{\pageref{firstpage}--\pageref{lastpage}}
\maketitle

\begin{abstract}
Ultra-short Period exoplanets (USPs) like 55\,Cnc\,e, hosting dayside magma oceans, present unique opportunities to study surface-atmosphere interactions. The composition of a vaporised mineral atmosphere enveloping the dayside is dictated by that of the surface magma ocean, which in turn is sensitive to its oxygen fugacity ($f$O$_2$). Observability estimations and characterisation of the atmospheric emission of 55\,Cnc\,e have mostly remained limited to low spectral resolution space-based studies. Here, we aim to examine ground-based high-resolution observabilities of a diverse set of mineral atmospheres produced across a grid of mantle $f$O$_2$s varying over 12 orders of magnitude. We assume a Bulk Silicate Earth mantle composition and a substellar dayside temperature of T = 2500K in the near infrared wavelength (NIR) region. This spectral range is often featureless for this class of atmospheres at low-resolution. Coupling our newly developed simulator for synthesising realistic observations from high-resolution ground-based spectrographs (\texttt{Ratri}) to a pre-developed high-resolution cross-correlation spectroscopy (HRCCS) analysis pipeline (\texttt{Upamana}), we find that this array of mineral atmospheres would all be detectable with 11 hours of observing time of the dayside of 55\,Cnc\,e with CARMENES and each individual scenario can be correctly differentiated within 1$\sigma$. Our analysis is readily able to distinguish between a planet with an Earth-like redox state (with $f$O$_2$ $\sim$3.5 log$_{10}$ units above the iron-wüstite, IW buffer) from a Mercury-like planet ($f$O$_2$ $\sim$5 log$_{10}$ units below IW). We thus conclude that the HRCCS technique holds promise for cataloguing the diversity of redox states among the rocky exoplanetary population.
\end{abstract}


%%%%%%%%%%%%%%%%%%%%%%%%%%%%%%%%%%%%%%%%%%%%%%%%%%

%%%%%%%%%%%%%%%%% BODY OF PAPER %%%%%%%%%%%%%%%%%%

\section{Introduction}\label{introduction}
\subsection{Ultra-Short Period Exoplanets} \label{USPs}
Ultra-Short Period exoplanets (USPs) are a class of exoplanets with periods $<$ 1 day and with radii $<$ 2 R$_{\oplus}$ \citep{sanchis2014study,winn2017absence,winn2018kepler,uzsoy2021radius}. The discovery of CoRoT-7\,b \citep{leger2009transiting,queloz2009corot} heralded this new class of exoplanets. Most exoplanets in this category with sizes below 1.5-1.6 R$_{\oplus}$ are thought to be predominantly rocky i.e. composed mostly of Fe and silicates \citep{weiss2014mass}, and exoplanets falling above that range are low density enough that Fe and silicates can no longer explain their radii and they must have a volatile envelope as well \citep{rogers2015most}. 
Planets either side this threshold could originate from
%Both possibilities could have been results of 
close-in Jovian sized gas giants that are stripped of their primary H/He rich atmosphere due to photoevaporation \citep[although the fraction of the observed USPs that could be formed through such a mechanism is low - see][]{winn2017absence}, or due to Roche lobe overflow \citep{valsecchi2014hot,valsecchi2015tidally,konigl2017origin} \citep[see][for a contrary result]{jackson2017new}. However, the scenario that best explains most observed USPs is that they are formed from smaller gaseous planets (Super-Earths and sub-Neptunes) in the initial size range of 2-4 R$_{\oplus}$ through photoevaporation \citep{owen2017evaporation,lundkvist2016hot,lopez2017born} and/or magnetospheric truncation of discs followed by tidal migration inward from the disc edge \citep{lee2017magnetospheric}. The study of this class of exoplanets existing in extreme environments provides an exciting opportunity to understand how exoplanets can orbit so close to their host stars and how they eventually evolve in such conditions.

\subsection{Lava/Magma Ocean Planets (LOPs/MOPs)} \label{lops}
%Extreme radiation, loss of their primary atmospheres and the fact that most of the USPs are expected to be tidally locked owing to the closeness to their host star means that the substellar temperature in the side permanently facing the star (dayside) can exceed 1500K. On the Earth's surface, deep below volcanic regions where such high temperatures can be reached, the silicate crust partially melts and flows \citep{chao2021lava} as 
%The extreme radiation from host stars can cause temperatures to reach $>$1500~K at the surface of tidally locked USPs. This is sufficient to exceed the 1 bar solidus for peridotite \citep[Earth composition][]{hirschmann2000mantle}. %Although there's no solar system planet analogue for the extreme conditions in which USPs currently exist, the Jovian tidally locked satellite Io is also driven to surface volcanism by the vigorous tidal forces exerted by Jupiter which heats up its interior \citep{peale1979melting,keane2023tidal}. 
USPs experience both %kinds of extreme situations of 
intense dayside heating as well as strong interior heating due to tidal forces from their host star \citep{jackson2008tidal}. This can cause the temperature to reach $>$1500~K at the dayside surface of tidally locked USPs which is sufficient to exceed the 1 bar solidus for peridotite \citep[Earth composition,][]{hirschmann2000mantle}. %which is why we also expect the silicate rocks on the dayside of USPs to undergo similar melting even though the initial composition of such rocks can be very different from those found on Earth, as well as the surface to have widespread violent volcanic activity. 
%These speculations have 
This has led to USPs being termed as Magma-Ocean Planets (MOPs) or Lava-Ocean Planets (LOPs) \citep{leger2011extreme} as their dayside is expected to be permanently melted to form a large scale magma ocean %or have large and deep pools/seas of lava (a local term for magma erupted from volcanoes)
\citep{chao2021lava}. The nightside of such LOPs would have low temperatures \citep{leger2011extreme} as it permanently faces away from the star. The temperature contrast between the hemispheres can then lead to (magma) oceanic circulations from the dayside to the nightside, however, it might not suffice to transport heat quickly enough for the nightside to heat up \citep{kite2016atmosphere}. %The depth of lava or magma seas/oceans on the dayside is governed by the balance of interior tidally induced heating and the radiation induced heating at the surface with minimal contribution from oceanic circulation \citep{kite2016atmosphere,lai2024oceancirculationtidelockedlava,herath2024thermal}, and is thus expected to become shallower as the system ages and the tidal forces dissipate which leads to cessation of the interior source of heating.
\\
\\
% \subsection{Mineral atmospheres in LOPs/MOPs - interior connection, temporal variation, and observability} \label{mineralatmospheres}
Apart from the extreme temperatures creating a magma ocean on the dayside and owing to the loss of most volatiles expected in primary atmospheres, the surface melt vaporises into a mineral (or dry) atmosphere \citep{schaefer2009chemistry,leger2011extreme,kite2016atmosphere}. This continues until the atmosphere comes into thermochemical equilibrium with the melt. Hence, the composition of the atmosphere is dictated by that of the underlying melt  \citep{schaefer2009chemistry}. Thus, retrieving the physical and chemical parameters of such vaporised atmospheres would help determine surface-atmospheric connections in USPs and act as a probe for melt composition. Melt compositions, in turn, reflect i) the bulk composition of the planet and ii) the pressure-temperature at its surface, % are affected by the interior compositions of the exoplanet, which are a result of the exoplanet's initial conditions and subsequent differentiation. Hence, probing the atmospheres of USPs also 
serving as an indirect window into the interior structure and composition of such exoplanets as well as possible markers to determine formation location in the initial protoplanetary disc. There is evidence that rocky planets, our Moon and planetesimals in the solar system could also have been covered by a magma ocean early during their formation \citep{wood1970lunar, fegley1987vaporization,chao2021lava}. Hence, studying USPs is also a possible analogue to studying our own solar system's past. 
\\
\\
% Attempts to provide a quantitative basis to the hypotheses outlined above arose from studies of the planet closest to the Sun in our own solar system, Mercury. One of the competing hypotheses to explain Mercury's formation is the vapourisation of its liquid mantle early in the solar system history when the temperature at its current location would exceed 2500K \citep{cameron1985partial}. Subsequently a model called \texttt{MAGMA} was developed to quantitatively describe such vapourisation leading to the high metal to silicate ratio on Mercury because the lighter silicates would be lost preferably over the heavier metals in such a vapourisation process \citep{fegley1987vaporization}. A modified version of \texttt{MAGMA} was also used to explain lava vapourisation in Io by \citet{schaefer2004thermodynamic}. 
Early studies of the composition of Mercury led to the development of the \texttt{MAGMA} code, which solves for the vapour-melt equilibrium at a magma ocean surface assuming a simplified, ideal liquid \citep{fegley1987vaporization, schaefer2004thermodynamic}.
The discovery of USPs subsequently provided the impetus to extend the use of \texttt{MAGMA} to that category of exoplanets and the initial studies were performed in \citet{schaefer2009chemistry} (for the case of CoRoT-7b alone) and \citet{miguel2011compositions} (for a larger grid of temperatures and distances). \citet{miguel2011compositions} found that the composition of vaporised atmospheres was dominated by refractory species such as Fe, Mg, Na, O, O$_2$ and SiO, thus coining the term \emph{mineral atmosphere} to describe them. 
%The \texttt{MAGMA} code used in the above-mentioned studies used the Ideal Mixing of Complex Components (IMCC) thermodynamic framework \citep{hastie1982alkali,hastie1982thermodynamic,hastie1984thermodynamic,hastie1985predictive,hastie1986predictive} to model the activity of the silicate melt, which in turn were calibrated using silicate melts with high-K$_2$O content \citep{ito2015theoretical,van2023lavatmos}. 
\citet{ito2015theoretical} subsequently improved the treatment of surface magma melt chemistry by using the \texttt{MELTS} thermodynamic framework \citep{ghiorso1995extrapolation}, which applies a more accurate activity model (required for the calculation of partial pressures of vapour species), but is limited to compositions with SiO$_2$ between $\sim$40 -- 75 wt.~\% and thus may not cover the anticipated range of exoplanetary mantle compositions \citep[e.g.,][]{spaargaren2023plausible}. Their approach was later adopted by \citet{van2023lavatmos}, \citet{charnoz2023effect}, and \citet{wolf2023vaporock} who also further included an appropriate treatment for the oxygen fugacity (\fOtwo) as mineral atmospheric composition strongly depends on it \citep{sossi2019evaporation, seidler2024impact}.
%for the silicate melt which has been calibrated experimentally on natural oxides with low-K$_2$O content to do a similar calculation for the most abundant species in such atmospheres and found a similar list (Na, K, Fe, Si, SiO, O, O$_2$). The \texttt{MELTS} formalism has also been used to construct two other codes to study outgassed/vapourised atmospheres on lava ocean exoplanets, namely \texttt{LavAtmos} \citep{van2023lavatmos} and \texttt{VapoRock} \citep{wolf2023vaporock}.
\\
\\
\citet{ito2015theoretical} were also the first to integrate their vapour-melt equilibrium chemistry model into a radiative transfer scheme in order to predict the observable physical and chemical properties of mineral atmospheres of USPs. They showed that atmospheric structure and radiation physics was predominantly shaped by SiO absorption in the UV and subsequent emission in the IR, a configuration that leads to strong thermal inversion, making such planets amenable to observation in emission/secondary eclipse observations. Follow-up studies by \citet{zilinskas2022observability}, \citet{piette2023rocky} and \citet{seidler2024impact} validated this finding and highlighted the importance of additional molecules, such as MgO and TiO, in shaping the emission spectrum. All these studies use models that assume a steady-state where the melt composition is held constant. In reality, magma ocean planets might also undergo chemical evolution \citep{kite2016atmosphere} through coupled magma ocean and atmospheric circulation \citep{nguyen2020modelling, nguyen2022impact}, cloud formation \citep{mahapatra2017cloud,nguyen2024clouds}, and atmospheric escape \citep{nguyen2020modelling,ito2021hydrodynamic}. 
\\
\\
\subsection{Effect of oxygen fugacity on mineral atmosphere composition} \label{fugeffect}
A key thermochemical quantity in any geochemical setting is the chemical potential of O$_2$ ($\mu_\mathrm{O_2}$). This quantity is set through the relation:% \citep{seidler2024impact}:
\begin{equation}
    \mu_\mathrm{O_2} = \mathrm{RT}\ln(f\mathrm{O_2}),
\end{equation}
where R is the universal gas constant, T the temperature of the system and \emph{f}O$_2$ is the oxygen fugacity, related to the partial pressure of O$_2$ via the fugacity coefficient $\Phi$: $f\text{O}_2 = \Phi_{\text{O}_2} p{\text{O}_2}$. %The state of fugacity is either based on metallic oxidation by H$_2$O \citep{elkins2008coreless} or by the oxidation state of the planetary building source material itself \citep{wang2019volatility} and has enormous implications on the core-mantle differentiation and size of the core. As an example specifically for iron (Fe), a highly oxidising environment would lead to a smaller core with most iron oxides (Wustite) going into the bulk mantle and a highly reducing environment would lead to a larger core with most metallic Fe residing in the core after differentiation \citep{wang2018,seidler2024impact}. Understanding the Wustite to Fe ratio is then one way to calculate \emph{f}O$_2$.
The $f$O$_2$ is a function of the pressure, temperature and composition of the system of interest. For planetary bodies, it can be calculated through the equilibrium relationship between iron (Fe) and wüstite (FeO), IW, set by:
\begin{equation}
\mathrm{Fe} + \frac{1}{2}\mathrm{O_2} \rightleftharpoons \mathrm{FeO}.
    \label{eq:IW}
\end{equation}
Hence, at equilibrium:
\begin{equation}
\mathrm{K} = \frac{a_{FeO}}{a_{Fe}.(fO_2)^{1/2}},
    \label{eq:IW_K}
\end{equation}
where K is the equilibrium constant, and $a_\mathrm{Fe}$ and $a_\mathrm{FeO}$ represent the activity of Fe in the metallic core and the activity of the FeO component in the multicomponent silicate mantle of the planet respectively. Thus, the mantle/core ratio is used as a proxy for the oxidation state of the planet \citep[e.g.,][]{righter2006compositional}.
For a USP with a tenuous mineral atmosphere where the bulk of the mass is in the melt i.e. the `vapour limited case' as defined in \citet{wolf2023vaporock}, %the oxidation state of the combined melt-atmosphere system is also controlled by 
the \emph{f}O$_2$ of the atmosphere, and hence the partial pressures of other vapour species, reflects the composition of the silicate liquid and the (irradiation) temperature. %and the atmospheric abundances are also expected to be strongly influenced by that quantity.
Because the compositions of silicate liquid are unknown for exoplanets, \citet{wolf2023vaporock} treated \emph{f}O$_2$ as a free parameter (in their vapour-melt equilibrium code \texttt{VapoRock}) and proposed that the  $p$SiO/$p$SiO$_2$ ratio could be used to determine the $f$O$_2$ of the atmosphere observationally. \citet{seidler2024impact} used the \texttt{MAGMA} code and also left $f$O$_2$ as a free variable, and to perform a large scale study %analogues to \citet{ito2015theoretical} and \citet{zilinskas2022observability} 
and determined that differences in silicate melt compositions, but predominantly in \emph{f}O$_2$ (varying across 12 orders of magnitude centred at IW $=$ 0), dictate the composition of the mineral atmosphere in equilibrium with the silicate liquid melt. They note that %where IW is the expected fugacity from the iron-wustite redox buffer at the particular temperature T) has a strong impact of the diversity of mineral atmospheres and showed that 
the line contrast ratio between the SiO and MgO features in emission in the mid-infrared (MIR) observed using JWST-MIRI could yield \emph{f}O$_2$ estimates to within $\pm$1 log$_{10}$ unit for 5 occultations of 55 Cnc e.

\subsection{The puzzling case of 55 Cnc e} \label{55cncepuzzle}
55\,Cnc\,e is the innermost planet in the 55\,Cancri (or 55\,Cnc) system and was discovered by \citet{mcarthur2004detection} %even before the discovery of the first USP. However, they 
who reported an orbital period of 2.808 days and M$\sin$i of 14.21$\pm$2.91 M$_{\oplus}$. This value was later discovered to be an alias of the actual lower orbital period of 0.7365 days and minimum mass of 8.3$\pm$0.3 M$_{\oplus}$ by \citet{dawson2010radial}. %, who also motivated looking for the expected hot planet, akin to CoRoT-7b, in transit. Following that, 
The exoplanet was successfully observed in transit at the predicted orbital period in the visible using MOST \citep{winn2011super} and in the infrared using Spitzer \citep{demory2011detection}. %Both studies also reiterated the fact that since the exoplanet transited a very bright naked eye K-type parent star (55\,Cnc\,A), it would be amenable to thorough atmospheric characterisation. 
From the mass and radius constraints obtained using both studies and the planetary composition models worked in \citet{valencia2010composition}, initially a thin envelope of supercritical water on a rocky Earth like core was the suggested model to explain the exoplanet. %This was in contrast to the case of CoRoT-7b which was theorised to be a lava ocean world right from the outset \citep{leger2009transiting,schaefer2009chemistry,leger2011extreme}. The difference stems from the diverse approaches to understand planet composition of USPs because \citet{valencia2010composition} had earlier argued for a thin supercritical water envelop on a rocky core composition model for CoRoT-7b as well through density fits, effectively arguing against the lava ocean world ocean scenario posited in \citet{schaefer2009chemistry}. These approaches would remain distinct for some more time. Meanwhile, 
Later, a purely silicate rich rock with no envelope \citep{gillon2012improved} or carbon rich rock \citep{madhusudhan2012possible} with no envelope were also seen to be compatible for 55\,Cnc\,e. However, the study by \citet{rogers2015most} necessitated that %an exoplanet the size of 
55\,Cnc\,e would need to have an atmosphere to explain its size. %of vapours so the models that allowed for an atmosphere were more preferable. 
%\\
%\\
\citet{demory2012detection} measured thermal emission from the exoplanet at 4.5$\mu$m using Spitzer and calculated a brightness temperature of 2360$\pm$300 K which %being close to the equilibrium temperature of $\sim$2400K \citep{demory2016variability} 
indicated either the presence of an atmosphere with low Bond albedo and poor heat transport from the dayside to the nightside, or that a layer of the atmosphere close to the equilibrium temperature due to thermal inversion in the atmosphere was being probed. %Later on in \citet{demory2016map} a longitudinal brightness map of the same exoplanet was prepared which showed the nightside brightness temperature to be 1380$\pm$400 K and the dayside hemisphere was calculated to have a temperature of 2700$\pm$270 K, a differential of $\sim$1300K. This reiterated the inefficient redistribution of heat result from \citet{demory2012detection}. However, the hot spot on the dayside was shifted 41\textdegree$\pm$12\textdegree eastward from the substellar point, which either indicated the presence of an optically thick atmosphere only on the dayside limiting heat recirculation to only one hemisphere, or a low viscosity magma flow on an atmosphere-less rock causing the hot spot drift. \citet{kite2016atmosphere} found that a magma flow would however still be insufficient to describe the temperature on night side, which was warmer than it would be if permanently faced away from the star, and would still require an atmosphere of non-silicate volatiles. 
\citet{demory2016variability} found variability in the thermal dayside emission at 4.5$\mu$m, with the brightness temperature varying between $\sim$1300-2800 K and the amplitude of variation in planetary radius being between 0.2-0.8M$_{\oplus}$, which would require a strong opacity source at that wavelength. %Volcanic plumes were thought to be responsible for the variation in thermal emission but the chemical composition of such plumes couldn't be determined. It was however suggested that sources like HCN, CO and CO$_2$ could be the cause of the high opacity. In addition, the case for the previous water rich atmosphere composition weakened (but wasn't ruled out) and the suggestion of an Earth like interior composition similar to the case of Io strengthened with the plumes being composed of CO$_2$, sulphur rich compounds and particulate silicate matter. The steady state atmosphere could also be composed of mineral atmospheres as suggested in \citet{miguel2011compositions} or \citet{leger2009transiting}, atmospheres similar to Venus \citep{schaefer2011atmospheric}. Alternatively, the case of a torus around the exoplanet composed of charged dust with CO contributing to the opacity was also not ruled out. \citet{angelo2017case} were able to reproduce the previous Spitzer results and posited a CO or N$_2$ rich atmosphere.
\\
\\
%However, t
The claim of detection of an atmosphere composed of H$_2$+He and HCN in \citet{tsiaras2016detection} using HST/WFC3 complicated matters. H$_2$ was not supposed to be present in USP atmospheres as the timescale of evaporation of primary atmospheres in USPs is quite short ($\sim$1Myr) \citep{valencia2010composition}, but there were claims that the energy limited formula used in \citet{valencia2010composition} could result in overestimation of the loss rate for USPs \citep{johnson2013molecular,angelo2017case}. Taking this into stride, \citet{hammond2017linking} performed GCM simulations to show that an atmosphere of 90\%H$_2$+10\%N$_2$ with trace amounts of CO and H$_2$O, and SiO clouds arising from a dayside magma ocean (the possibility of which had been posited earlier that year by \citet{mahapatra2017cloud}) could account for the Spitzer phase curve observations in \citet{demory2016map}. They were also the first to propose a nightside lava ocean to explain the higher measured temperature than expected. %This was followed by a significant reassessment of the exoplanet's properties using data from HST/STIS in 
However, \citet{bourrier201855} %which 
ruled out a H/He dominated scenario or a supercritical water rich envelope in favour of a heavyweight envelope using data from HST/STIS. But, the study was unable to distinguish between the kinds of heavyweight envelope - a mineral atmosphere vaporised from a lava ocean case or a high mean molecular atmosphere dominated by either CO or N$_2$ case. Since then, these two options have been the competing hypotheses to explain the atmospheric composition of this exoplanet. %Consequently, the observability using JWST for 55\,Cnc\,e assuming a N$_2$ dominated atmosphere was then looked at by \citet{zilinskas2020atmospheric} and \citet{zilinskas2021temperature}, while the observability of mineral dominated atmospheres was done in \citet{zilinskas2022observability} and most recently in \citet{seidler2024impact}. 
\\
\\
%Meanwhile The observed 
The variability in emission continued to be present in the infrared %(reanalysis of Spitzer data in \citet
\citep{tamburo2018confirming}, weakly in the optical using MOST \citep{sulis2019multi}, TESS \citep{valdes2022weak} and using CHEOPS \citep{demory202355}. The CHEOPS study also deemed unlikely that the variation could be caused due to reflection from the exoplanet or due to excess emission from the star due to magnetospeheric interaction between the star and planet theorised in \citet{folsom2020circumstellar}. The most recent re-reduction and reanalysis of Spitzer data in \citet{mercier2022revisiting} however showed a much higher dayside temperature ($\sim$ 3771K), a higher differential between the dayside and nightside ($>$2122K), % and a negligible hotspot drift from the substellar point compared to the results in \citet{demory2016map}, and hence 
and favoured a mineral atmosphere composed of SiO atop a lava ocean case with inefficient atmospheric circulation. Hence, the tide between which heavyweight atmospheric scenario would be more appropriate keeps shifting over the years. %Even 
Recently, %when 
it finally seemed like a secondary C-rich atmosphere (with evidence of CO or CO$_2$) %which could be 
vaporised and sustained by a magma ocean was finally detected with a level of certainty using NIRCam and MIRI on JWST by \citet{hu2024secondary}. %, 
But, the results of an emission variability study by \citet{patel2024jwst}, %where they investigated the possibility of variation of emission caused due to asynchronous rotation due to the exoplanet being in a 3:2 resonance, 
indicated that not only are the variability across 2.1$\mu$m and 4.5$\mu$m channels not correlated, the atmospheres change from being described by CO/CO$_2$ rich to silicate mineral vapour to no atmosphere depending on the time of their observation. The puzzling nature of this exoplanet thus still continues for the time being.

\subsection{55 Cnc e in high resolution studies over the years} \label{55cncehighres}
The field of high resolution spectroscopy (R=$\lambda/\Delta\lambda >$25000) using ground based spectrographs has been used to investigate the nature of Jovian exoplanets successfully over the past decade, especially using cross-correlation spectroscopy (HRCCS) \citep[][to name a few]{brogi2012signature,birkby2013detection,de2013detection,de2014identifying,snellen2008ground,snellen2010orbital,giacobbe2021five} and in this decade has also been used to investigate the nature of sub-Neptunes \citep{lafarga2023hot,dash2024constraints}. %Hence, 
Investigating the nature of hot terrestrial super-Earths has also become an attractive option and %for such investigations and many studies pretty much complement space based investigations of the same exoplanets, and 
55\,Cnc\,e being a bright source around a bright host star provides a natural target. %an attractive option to start with. since it would be easy to get enough S/N to enter the photon noise dominated regime. 
\\
\\
The issue of whether the exoplanet has a mineral atmosphere or a high mean molecular weight atmosphere of N$_2$ or CO$_2$ remains complicated by the fact that there had been no strong detection of any molecular or atomic species in the exoplanet's atmosphere using ground based high resolution spectrographs as well. Non detection of escaping H in Ly$\alpha$ \citep{ehrenreich2012hint} (using HST/STIS) or He \citep{zhang2021no} (using NIRSPEC at Keck) indicates a non-existence of a low mean molecular weight atmosphere. \citet{ridden2016search} using data from 4 transits using HARPS found a weak signature of Na in the exosphere, along with a signature of Ca$^{+}$ in one of the transits %. However, 
but the signals were not significant enough to claim unambiguous detection. %, but escaping Na in the exosphere could be in line with what is expected for mineral atmospheres in USPs (see Section 1.3). 
%Another 
\citet{esteves2017search} performed a high resolution transmission study in the optical using HDS on the Subaru telescope and ESPaDOnS on the Canada-France-Hawaii Telescope (CFHT) and %also 
found no signature of H$_2$O. %\citep{esteves2017search}. , which ruled out the optically thin supercritical water envelope case proposed in \citet{demory2011detection}, \citet{winn2011super} and \citet{gillon2012improved}, and in support of results from HST/STIS from \citet{bourrier201855}. 
\citet{jindal2020characterization} used an additional four nights aside from the transmission datasets used in \citet{esteves2017search} and still obtained no signature of H$_2$O or TiO. Their results were consistent with a high mean molecular weight atmosphere, a cloudy atmosphere and even a no atmosphere case. \citet{tabernero2020horus} also performed a transmission study using HORuS at the Gran Telescopio Canarias but failed to replicate the NaI results from \citet{ridden2016search} %but were able to provide improved upper limit constraints. They 
and also did not detect any H$\alpha$. Motivated by results in \citet{tsiaras2016detection}, and the observational modelling for N dominated atmospheres in \citet{zilinskas2020atmospheric}, 
\citet{deibert2021near} performed a high resolution transmission study in the infrared using CARMENES at Calar Alto and SPIRou at CFHT but failed to detect an atmosphere. %, or any evidence of HCN, NH$_3$, C$_2$H$_2$, CO, CO$_2$ and H$_2$O which indicated either the presence of clouds or an even more compact high mean molecular weight atmosphere. 
\citet{keles2022pepsi} then undertook a high resolution transmission study in the optical to search for a mineral atmosphere %formed from rock vaporisation 
using the PEPSI instrument at the Large Binocular Telescope, but were unable to detect it. %. However, they found no evidence for Fe, Fe$^{+}$, Ca, C$^{+}$, Mg, K, Al, Ti, Ti$^{+}$, Mn, Mn$^{+}$, Ba, Ba$^{+}$, Sr, S, Zr, Zr$^{+}$, V and Cr. %which are major elements in the Earth's crust and could be present in a mineral atmosphere. However, since the exoplanet has a high brightness temperature, looking at it in emission is a better option to probe any kind of high mean molecular weight atmospheres including mineral ones. Hence, bolstered by a renewed interest in mineral atmospheres on this exoplanet due to the results from \citet{mercier2022revisiting} led to 
\citet{rasmussen2023nondetection} performed the first high resolution emission study using MAROON-X, but found no evidence for Fe.

\subsection{Placing this study in context} \label{resincontext}
All but one \citep[i.e.][]{rasmussen2023nondetection} previous studies at high resolution have been in transmission. An %non-uniformly thick 
atmosphere majorly restricted to the dayside %and potentially very thin and cloudy \citep{mahapatra2017cloud} at the terminators 
presents unique challenges for observations in transmission. Hence, emission spectroscopy is a potentially better avenue to try and characterise this exoplanet with its high Emission Spectroscopy Metric (ESM) value \citep{kempton2014high,seidler2024impact}, something which has now been tried with space based observations \citep{hu2024secondary,patel2024jwst}. Both studies showcase the unique challenges in trying to characterise 55\,Cnc\,e as it showcases temporal variation in the kind of atmosphere it fits for - sometimes a mineral atmosphere and sometimes a vapour rich atmosphere (or even a bare rock in some cases).%, but both atmospheric cases nevertheless necessitate the presence of a dayside surface magma ocean. 
The difference between both regimes is in the vapour content of the surface melt which can be additionally outgassed to form a mixed mineral atmosphere where in addition to the species present in the dry mineral atmosphere case we concentrate on in this study, other species like CO, CO$_2$ and H$_2$O can also be present. Hence, the use of interior or surface melt models to understand the kinds of atmospheres that could be possible on USPs like 55\,Cnc\,e is now overall a crucial step. \citet{keles2022pepsi} and \citet{rasmussen2023nondetection} both took into account the effect of the surface by assuming a Bulk Silicate Earth (BSE) composition for the mantle. \citet{keles2022pepsi} assumed a solid surface and used the elemental abundances corresponding to a BSE to model their isothermal atmospheres at the exoplanet terminators. \citet{rasmussen2023nondetection} used two kinds of models: 1. using BSE elemental abundances but with varying fractions of Fe enveloped by a Venusian CO$_2$ dominated atmosphere, and 2. with an assumption of magma-atmosphere interaction, albeit indirectly, where the BSE mantle was modelled using the \texttt{MELTS} framework \citep{ghiorso1995extrapolation}. In case 2, the vertical atmospheric abundances were not calculated and instead the atmosphere was assumed to be chemically evenly mixed at the abundances calculated at the magma-atmosphere interaction on the dayside of the exoplanet. These abundances were then  used to further post process a P-T profile generated using HELIOS \citep{malik2017helios}. Thus, there have not been any studies at high-resolution yet which have utilised a full coupled interior-atmosphere framework to model both the atmospheric P-T profile and the vertical chemical abundances, or have looked at an expanded grid of possible atmospheric compositions due to a variation of \emph{f}O$_2$ value of the mantle. As \citet{seidler2024impact} have shown, the composition (and structure) of atmospheres is a lot more sensitive to mantle \emph{f}O$_2$ compared to just the composition of the mantle assumed.
\\
\\
%\citet{piette2023rocky} recently presented the case for the observability of mixed vapour atmospheres using JWST-MIRI by using \texttt{VapoRock} \citep{wolf2023vaporock} for modelling the surface melt but also additionally incorporating volatiles into the mineral atmospheres and observability for mineral atmospheric cases have been presented in \citet{zilinskas2022observability} and \citet{seidler2024impact}. However, simulation studies showcasing observabilities/detectabilities for both atmospheric regimes in high-resolution have lagged behind. This study is an 
Here, we attempt to remedy this by computing a grid of mineral atmospheres generated across 12 orders of magnitude in \emph{f}O$_2$, assuming a BSE-like mantle composition, following the prescription laid down in \citet{seidler2024impact} for calculating relative abundances, partial pressures and temperatures. %and using two substellar temperature estimates corresponding to dilution factors ($f$) of $2/3$ (T = 2500K) and 1 (T = 3000K). We concentrate on the more realistic $f=2/3$ case in the main text and showcase the results for $f=1$ in Appendix B. On this grid of atmospheres, we determine relative abundances, partial pressures and temperatures of prevailing gas species following \cite{seidler2024impact}. %computed at high spectral resolution. 
We then focus on the near-infrared (NIR) window (0.8--2.5$\mu$m) since many ground-based high-resolution spectrographs are devised to work in this wavelength range. %Mineral atmospheres as modelled in \citet{seidler2024impact} are featureless at low-resolution, which hampers characterisation. Hence, opening up a potentially new window to characterise mineral atmospheres using this window will allow a wider avenue to characterise atmospheres of USPs more precisely using both space-based low-resolution and ground-based high-resolution data. 
To do so, %For the purpose of assessing detectabilities in the NIR, we first describe the process of building 
we develop our own %synthetic signal-to-noise ratio (SNR) 
pipeline \texttt{Ratri}\footnote{named after the Vedic goddess of night - R\=atri} to generate realistic high-resolution observations from some of the best performing spectrographs used in that wavelength region today - CARMENES, GIANO, SPIRou and CRIRES+, while also considering the time variation of telluric absorption through the Earth's atmosphere. Then we couple \texttt{Ratri} %this SNR pipeline 
to the HRCCS detrending and analysis pipeline we introduced in \citet{dash2024constraints} (here on named \texttt{Upamana}\footnote{Up\=amana means knowledge through analogy or comparison in Sanskrit}) which uses a Singular Value Decomposition/Principal Component Analysis (SVD/PCA) based detrending procedure and employs a robust Bayesian parameter estimation framework. As a result, we now have an an end-to-end simulator %(\texttt{Rupamana}\footnote{a shortening of Ratri$+$Upamana}) 
to check for detectabilities/observabilities of any exoplanet with known ephemerides, as described in Section \ref{method}. In Section \ref{results} the simulated nights and the degree to which mineral atmospheres at different $f$O$_2$ can be detected and distinguished with CARMENES is discussed. %the detectabilities and differentiabilities of all the diverse atmospheric cases utilising our HRCCS pipeline. 
Section \ref{discussion} lists the utility of our framework for revealing the internal state of rocky exoplanets and potential caveats, before we conclude in Section \ref{conclusion}. %This study is a high-resolution follow-up to the work done in \citet{seidler2024impact} and through its results aims to motivate future observations of 55\,Cnc\,e using ground-based high-resolution spectrographs in emission. %to aid in decoding the continuing mystery regarding the atmospheric composition of this exoplanet.

\begin{table*}
    \centering
    {\renewcommand{\arraystretch}{1.5}
    \begin{tabular}{|c|c|c|c|}
    \hline
    \textbf{Parameter} & \textbf{Description} & \textbf{Value} & \textbf{Reference} \\
    \hline
    \hline
    M$_{\star}$ & Stellar mass & $0.905^{+0.015}_{-0.015}$ M$_{\odot}$ & \citealt{bourrier2018hubble} \\
    R$_{\star}$ & Stellar radius & $0.943^{+0.010}_{-0.010}$ $R_{\odot}$ & \citealt{bourrier2018hubble} \\
    T$_\mathrm{eff}$ & Effective temperature & $5172^{+0.010}_{-0.010}$ K & \citealt{bourrier2018hubble} \\
    \hline
    M$_\mathrm{P}$ & Planet Mass & $7.99^{+0.32}_{-0.33}$ M$_{\oplus}$ & \citealt{bourrier2018hubble} \\
    R$_\mathrm{P}$ & Planet Radius & $1.875^{+0.029}_{-0.029}$ R$_{\oplus}$ & \citealt{bourrier2018hubble} \\
    $a$ & Semi-major axis & $3.52^{+0.01}_{-0.01}$ $R_{\star}$ & \citealt{bourrier2018hubble} \\
    $P_\mathrm{P}$ & Orbital Period & $0.7365474^{+0.00000131}_{-0.00000141}$ days & \citealt{bourrier2018hubble} \\
    T$_\mathrm{0}$ & Transit Mid-Point & $2457063.2096^{+0.0006}_{-0.0004}$ BJD & \citealt{bourrier2018hubble}\\
    T$_{14}$ & Transit Duration & $1.58^{+0.08}_{-0.07}$ hours & \citealt{bourrier2014detecting} \\
    $e$ & Eccentricity & $0.05^{+0.03}_{-0.03}$ & \citealt{bourrier2018hubble} \\
    $v_\mathrm{sys}$ & Systemic Velocity & $27.26^{+0.14}_{-0.14}$ km\,s$^{-1}$ & \citealt{brown2018gaia} \\
    $i$ & Orbital Inclination & 89.59$^{\circ +0.47^{\circ}}_{-0.44^{\circ}}$ & \citealt{bourrier2018hubble} \\
    $K_\mathrm{P}$ & Exoplanet Semi-amplitude velocity & $228.49^{+2.52}_{-2.52}$ km\,s$^{-1}$ & Expected value, this study \\
    \hline
    \end{tabular}
    }
    \caption{Parameters for 55\,Cnc\,A (host star) and 55\,Cnc\,e (exoplanet) used in this study and their values}
    \label{tab:system_par}
\end{table*}

\section{Methodology}\label{method}
In this section, we first describe the creation and workflow of our custom SNR pipeline, which we then use to generate the synthetic nights of observations for our HRCCS analysis. Then we briefly describe our custom detrending pipeline for HRCCS analysis, since \citet{dash2024constraints} already provides a detailed description of the same workflow. We follow by finally describing the generation of abundance and P-T profiles for the diverse atmospheres possible due to changes in oxygen fugacity imposed on the initial mantle composition of 55 Cnc e and the subsequent generation of spectra using the GENESIS pipeline \citep{gandhi2017genesis}.
\subsection{Signal-to-Noise Ratio (SNR) pipeline}\label{pipeline}
\subsubsection{Stellar Spectrum and path to top of Earth's atmosphere}\label{phoenix}
We use a PHOENIX spectrum with the same parameters as 55 Cnc A as shown in Table \ref{tab:system_par}. This is downloaded from the High Resolution Spectra archive\footnote{ftp://phoenix.astro.physik.uni-goettingen.de/HiResFITS/} maintained at the Gottingen Spectral Library\footnote{https://phoenix.astro.physik.uni-goettingen.de} \citep{husser2013new}. We note here that our calculator itself doesn't mandate the use of a stellar spectrum from any specific catalogue, as long as the resolution of the obtained spectrum is $>$100,000 and the wavelength range (mentioned below) is sufficient for our requirements.
\\
\\
We limit the PHOENIX spectrum to cover a wavelength range of 0.35-2.53$\mu$m. This covers the wavelength ranges of CARMENES, GIANO, SPIRou, and covers the Y, J, H and K bands of CRIRES+, which are the current generation ground based spectrographs we consider in this study. We restrict the band of CRIRES+ to K because the thermal background ramps up in the L and M bands and we do not account for this increase in our calculator at the moment. Our calculator is designed for use in cases where the targeted star is the source of most of the photons received by the detector (photon noise dominated regime). We first broaden the spectrum using a Gaussian kernel down to the resolution of the instrument we need to consider for our simulation, perform a spline fit and save the fitted parameters for use later, and then convert the broadened spectrum obtained in units of flux density ($F_{\lambda,\star}$) to units of luminosity ($L_{\lambda,\star}$) through the relation
\begin{equation}
    L_{\lambda,\star} = 4 \pi R_{\star}^{2} F_{\lambda,\star},
\end{equation}
where $R_{\star}$ is the stellar radius of 55 Cnc A. We then account for the path of light to the top of Earth's atmosphere and find the corresponding flux density received there ($F_{\lambda, \mathrm{etop}}$) through the relation
\begin{equation}
    F_{\lambda, \mathrm{etop}} = L_{\lambda,\star}/ 4 \pi d^{2},
\end{equation}
where $d$ is the distance to 55 Cnc A. Then we convert this flux density to units of photon flux density ($N_{\lambda, \mathrm{etop}}$) by dividing through by the corresponding wavelength dependent energy of photon quanta ($hc/\lambda$). Then we spline fit this spectrum and save the relevant fit parameters for use later.  

\subsubsection{Telluric Absorption on path to detector}\label{telluric}
Next we need to take the path of light through the Earth's atmosphere into consideration. For this, we measure the effect of telluric absorption in the Earth's atmosphere ($T_{\lambda,\oplus}$) by obtaining a telluric model of the Earth's atmosphere at a resolution of 200,000 through SkyCalc CLI\footnote{https://www.eso.org/observing/etc/doc/skycalc/helpskycalccli.html}, which is a command-line interface of the ESO Sky Model Calculator \citep{noll2012atmospheric,jones2013advanced}. Generation of each telluric spectrum generally needs a few seconds and hence to speed up our end-to-end simulations, we first create a repository of telluric spectra for all possible Precipitable Water Vapour (PWV) values supported by the CLI with the airmass value fixed at 1.0. During the actual simulation, we only load in the spectrum from this repository corresponding to the PWV used per spectrum. What distribution of PWVs is used depends on the quality of night. We designate PWVs of 2.5 and 3.5 mm for `Good' nights, 5 mm and 7.5 mm for `Average' nights and 10 mm for `Bad' nights. Additionally, to account for change in this transmission model due to airmass variations, either for using different values in our simulations or to account for time variation during simulated nights of observations ($\mu(t)$), we use the relation
\begin{equation}
    T_{\lambda,\oplus}(\mu(t)) = e^{[\mu(t)\mathrm{ln}(T_{\lambda,\oplus}(\mu = 1))]}.
\end{equation}
This relation follows from the Beer-Lambert (or Bouguer-Lambert-Beer) law \citep{bouguer1729essai,lambert1760photometria,beer1852bestimmung}. The effect of absorption through the Earth's atmosphere serves as an extra multiplicative component to the flux passing through the atmosphere to reach the spectrograph's detectors. 
\subsubsection{Detector response and SNR output}\label{detector}
Since we are eventually concerned with the detector response to the incident photon flux, we first evaluate the values of incident photon flux at the top of Earth's atmosphere and the telluric model absorption values at the wavelength grid for each spectrograph's detector using the spline fit parameters for these saved in Section \ref{phoenix} and Section \ref{telluric} respectively. The wavelength grids for CARMENES, GIANO and SPIRou are based on values computed from archival observations previously used for a similar purpose as our work in \citet{gandhi2020seeing}. For CRIRES+, we extracted the wavelength grids for each instrumental mode from the S/N per spectral pixel output generated by command-line querying\footnote{through the etc\_cli.py script and a template input JSON file provided in the online version of the ESO ETC 2.0 calculator} of the ESO ETC 2.0 calculator \citep{boffin2020eso} using stellar parameters corresponding to 55 Cnc A, with the Adaptive Optics (AO) system enabled at the default values, default values for the Sky model parameters (which has airmass $=$ 1.2 and PWV $=$ 2.5 mm) and integration time parametrs of NDIT$=$1 and DIT$=$30s. We additionally also query for the total throughput efficiencies and atmospheric transmission and save them for later use.
\\
\\
The wavelength grid of the detector is generally a 2D array of size ($n_\mathrm{orders}\times n_\mathrm{pixels}$) because each spectrograph considered here is a cross-dispersed spectrograph where a 2D spectrum with $n_\mathrm{orders}$ spectral orders is produced on CCDs spanning $n_\mathrm{pixels}$ in the horizontal dimension. Hence, after traversing through the Earth's atmosphere the photon flux received on the detector CCDs is given as
\begin{equation}
    N_\mathrm{res, detector} = T_{\mathrm{res},\oplus}(\mu(t)) N_\mathrm{res, etop},
\end{equation}
where $N_\mathrm{res}$ and $T_{\mathrm{res}}$ are the photon flux and telluric model absorption respectively evaluated at the resolution elements of the detector CCDs.
\\
\\
While the previous relation gives the photon flux reaching the detector CCDs, the number of photons collected at each resolution element (N$_\mathrm{tot,res,detector}$) of the detector is what we are interested in since we need to calculate SNR. To model this, we use the effective collecting area of the telescope on which the spectrograph considered in our simulation is mounted ($A_\mathrm{spec}$), the integration time for collection of photons ($t_\mathrm{int}$) and the width of each resolution element of the detector ($w_\mathrm{res}$) to calculate the number of photons collected at the detector CCDs. However, not all incident photons are recorded on the detector and this measure is provided by the instrumental throughput ($\epsilon_\mathrm{res}$) for each resolution element, which also includes the quantum efficiency of the detector. Since each of these components serves as a multiplicative factor to the photon flux, the total number of photons collected at each resolution element is simply
\begin{equation}
    \mathrm{N}_\mathrm{tot,res,detector} = A_\mathrm{spec}  t_\mathrm{int}  w_\mathrm{res}  \epsilon_\mathrm{res}  N_\mathrm{res, detector}.
\end{equation}
As before, the $\epsilon_\mathrm{res}$ values for CARMENES, GIANO and SPIRou were obtained from archival observations. For CRIRES+, we manually calculate the $\epsilon_\mathrm{res}$ grid by dividing the total throughput efficiency output by the atmospheric transmission output for each instrumental mode (both obtained as mentioned earlier from command-line querying). The $w_\mathrm{res}$ grid for all spectrographs is calculated from their wavelength grids by subtracting consecutive wavelength values (higher$-$lower) for each spectral order. Assuming enough integration time where Poisson statistics approaches that of a Normal distribution, the photon noise/flux error term approaches $\sqrt{\mathrm{N}_\mathrm{tot,res,detector}}$ and hence the SNR values we need are obtained as
\begin{equation}
    \mathrm{SNR}_\mathrm{res,detector} = \frac{\mathrm{N}_\mathrm{tot,res,detector}}{\sqrt{\mathrm{N}_\mathrm{tot,res,detector}}}.
\end{equation}

\subsection{Simulating nights of observations using \texttt{Ratri}} \label{nightsimul}
\begin{table*}
    \centering
    \begin{tabular}{|c|c|c|c|c|}
    \hline
    \textbf{Spectrograph} & \textbf{Resolution} (R) & \textbf{t$_\mathrm{overhead}$ (with no nodding) (s)} & \textbf{t$_\mathrm{nodding}$ (s)} & \textbf{Area of the aperture} (m$^2$) \\
    \hline
    \hline
    CARMENES & 80400 & 34 & - & 9.00 \\
    GIANO & 50000 & 60 & - & 9.45 \\
    SPIRou & 70000 & 29 & - & 8.17 \\
    CRIRES+ & 86000 (0.2'' slit) & 360 (preset), 240 (acquisition with AO), 2.4 (RO with NDIT = 1) & 28 & 52.81 \\
    \hline
    \end{tabular}
    \caption{Parameters of the spectrographs used in this study. The values for CARMENES, GIANO and SPIRou are taken from \citet{gandhi2020seeing}. Parameters for CRIRES+ are from the ESO user manual for P114 Phase 2. The aperture area is based on the usage of an 8.2\,m diameter mirror at the VLT.}
    \label{tab:spec_par}
\end{table*}
\subsubsection{Scheduling observations} \label{scheduleobserve}
Our pipeline \texttt{Ratri} starts off by generating a list of observation times around a few (user specified) hours of a date and time provided in Coordinated Universal Time (UTC). The resolutions, duty (and nodding) times, and effective aperture area for telescopes on which the spectrographs considered in this study are mounted are shown in Table \ref{tab:spec_par}. We use the \texttt{astroplan} package \citep{morris2018astroplan} to generate airmass and altitude values of the 55 Cnc system at the observation times thus obtained, corresponding to n$_\mathrm{spectra}$ number of exposures being taken over the entire duration of observation. Then we use the \texttt{astropy} package to %convert the times to first Barycentric Dynamical Time (TDB) and then to Julian Dates (JD). Then we convert them to Barycentric Julian Date (BJD) format as well by accounting for the correction due to the time it takes for light to travel from the target to Earth
convert the UTC observation times (in JD$_\mathrm{UTC}$) to JD$_\mathrm{TDB}$, where JD is the Julian Date and TDB is Barycentric Dynamical Time. The JD$_\mathrm{TDB}$s are then converted into Barycentric Julian Dates (BJD$_\mathrm{TDB}$, or simply BJD from here on) by accounting for the correction due to the time it takes for light to travel from the target to Earth. We also generate Barycentric Radial Velocity corrections for each observation time (in km\,s$^{-1}$). The generated BJD values and Barycentric Radial velocity values are saved for later use.

\subsubsection{Accounting for time varying spectrum from the orbiting exoplanet} \label{doppler}
The pre-calculated Barycentric Radial Velocity corrections are first used to calculate the planetary orbital velocities ($v_\mathrm{P}(t)$) through the relation
\begin{equation}
    v_\mathrm{P}(t) = v_\mathrm{sys}-v_\mathrm{bary}(t)+K_\mathrm{P}\{ \cos[f(t)+\omega] + e\cos(\omega)\},
\end{equation}
where $v_\mathrm{bary}(t)$ are the barycentric radial velocities of the observer throughout the observation, which were saved in Section \ref{scheduleobserve}. $\omega$ is the longitude of periastron and $e$ is the eccentricity of the exoplanet's orbit and we use the values provided in Table \ref{tab:system_par}. $ v_\mathrm{sys}$ is the systemic velocity of the system and is also taken from Table \ref{tab:system_par}. We note that for the exoplanet we simulate in this study, 55\,Cnc\,e, the value of $e$ is compatible with 0 (or a circular orbit) at 1$\sigma$. However, we still use the precise value just to show that our simulator is capable of handling eccentric orbits as well. $K_\mathrm{P}$ is the exoplanet Semi-amplitude velocity and is calculated as
\begin{equation}
    K_\mathrm{P} = \frac{2\pi a}{P_\mathrm{P}\sqrt{1-e^{2}}} \sin(i),
\end{equation}
where $a$ is the semi-major axis, $P_\mathrm{P}$ is the orbital period and $i$ is the orbital inclination. $f(t)$ in Equation 7 is the true eccentric anomaly, which is calculated as:
\begin{equation}
    f(t) = 2\arctan\bigg{(}\frac{\sqrt{1+e}}{\sqrt{1-e}}\bigg{)}\tan\bigg{(}\frac{E(t)}{2}\bigg{)}.
\end{equation}
$E(t)$ is the Eccentric anomaly, which is obtained by solving the Kepler equation connecting it to the Mean anomaly ($M(t)$) numerically:
\begin{equation}
    E(t) - e\sin[E(t)] - M(t) = 0.
\end{equation}
The time varying $M(t)$ is found through its relation to the exoplanet orbital phases:
\begin{equation}
    M(t) = 2\pi\phi(t) - \pi/2 -\omega.
\end{equation}
The exoplanet orbital phases ($\phi(t)$) are calculated from the BJDs pre-calculated and saved in Section \ref{scheduleobserve}.
\\
\\
Our considered exoplanet spectrum generated through any radiative transfer code (GENESIS for this study, see Section \ref{genesis}) is first regridded and broadened to the resolution of the spectrograph used for the simulation. For each observation time calculated in Section \ref{scheduleobserve}, the regridded and broadened spectrum is Doppler shifted through the velocity calculated in Equation 10 and then spline fitted and the fitting parameters saved. These fitting parameters are used to evaluate the exoplanet spectrum on the wavelength grid of the PHOENIX stellar spectrum saved in Section \ref{phoenix}. The resultant stellar$+$exoplanet spectrum is calculated by multiplying this time varying additional contribution of the exoplanet. For an exoplanet emission spectrum contribution, the additional contribution is provided as the ratio $F_{\lambda, \mathrm{P}}/F_{\lambda,\star}$ which are loaded in from the model template generated through any radiative transfer code. Thus the combined spectrum ($F_{\lambda,\star+\mathrm{P}}$) is calculated as:
\begin{equation}
    F_{\lambda,\star+\mathrm{P}} = F_{\lambda,\star}\bigg{(}1+\frac{F_{\lambda, \mathrm{P}}}{F_{\lambda,\star}}\bigg{)}.
\end{equation}
Although not covered in this study, \texttt{Ratri} can also work when the input is a transmission spectrum output from any radiative transfer code %from GENESIS, 
provided in $(R_{\lambda, \mathrm{P}}/R_{\star})^{2}$ units, which is the wavelength dependant transit depth of the planet. $R_{\lambda, \mathrm{P}}$ is the notation for the wavelength dependant radius of the exoplanet. %we also provide the calculation where a transmission spectrum is used for completeness. %For an exoplanet transmission spectrum, 
%In that scenario, the contribution that is absorbed by the exoplanet's atmosphere (and hence subtracted from the stellar flux) is proportional to the area blocked out by the exoplanet at each wavelength and is calculated as $(R_{\lambda, \mathrm{P}}/R_{\star})^{2}$, which is the transit depth of the planet at each wavelength ($R_{\lambda, \mathrm{P}}$ is the notation for the wavelength dependant radius of the exoplanet). The values of transit depth are loaded in from the template exoplanet model generated using GENESIS. Hence 
The combined spectrum in such a case is calculated as:
\begin{equation}
    F_{\lambda,\star+\mathrm{P}} = F_{\lambda,\star}\bigg{[}1-\bigg{(}\frac{R_{\lambda, \mathrm{P}}}{R_{\star}}\bigg{)}^{2}\bigg{]}. 
\end{equation}
$F_{\lambda,\star+\mathrm{P}}$ is then spline fitted and saved. The procedure to calculate the number of photons at the detector for this combined case is the same as outlined in Section \ref{pipeline}. Since each observation time has had its airmass value calculated in Section \ref{scheduleobserve}, this is the airmass value we use in Equation 6, along with a value for PWV randomly selected between the values allowed based on the quality of night, to calculate the time varying telluric absorption model. An ability to incorporate time variance of telluric absorption through the Earth's atmosphere introduces an additional source of complexity to our end to end simulator \texttt{Ratri}, not considered in \citet{gandhi2020seeing}.%, which simply used a single value for each of PWV and airmass for their simulations.

\subsubsection{Addition of white noise and instrumental variation effect} \label{noise}
For each of the exposures generated at the end of the workflow depicted in Section \ref{doppler}, we perform all the steps as listed in Section \ref{pipeline}. This gives us photon flux, photon flux error and wavelength arrays of size ($n_\mathrm{orders}\times n_\mathrm{pixels}$) for each of the $n_\mathrm{spectra}$ exposures taken during the night of observation, which can then all be rearranged into ($n_\mathrm{orders}\times n_\mathrm{spectra}\times n_\mathrm{pixels}$) sized cuboids. Assuming that the errors at all pixel are uncorrelated, the actual noise at each pixel can be randomly drawn from a normal distribution with variance equivalent to the photon flux error value calculated at that pixel assuming Poisson statistics. To incorporate this, we generate an additive white noise component at each pixel by picking random values from a normal distribution centred at 0 and with a variance equivalent to the photon flux error value calculated at that pixel. This white noise is added to the photon flux received by adding together both cuboids. A very small fraction of the pixels are found to have a net photon flux value below 1. The values of these pixels are all changed to 1 to make sure that they will be masked through the \emph{low flux masking} step we perform while preprocessing our cuboids and before detrending in Section \ref{pca}. To account for pointing inaccuracy and variation in overall atmospheric transparency, we multiply the photon fluxes at each exposure by a number picked randomly from a uniform list containing values between 0.95 to 1.05 to account for a variation of at most $\pm$5\% in the photon flux actually received.
\\
\\
We henceforth label the resultant white noise and instrumental effect incorporated photon flux cuboid as \textbf{A}, and the white noise flux error cuboid as \textbf{$\varepsilon$} to keep our notations in line with that of \citet{dash2024constraints}. These cuboids, alongside the values of the phases and barycentric radial velocities (each of size $n_\mathrm{spectra}$) already pre-generated, are used for our next step of trying to detect the exoplanet signal embedded in the flux cuboid through the use of an SVD/PCA based detrending procedure followed by cross-correlation with the template model signal.
\\
\\
During real observations, strong correlated noise (red noise) can arise for narrow spectral features due to PWV variations and/or due to small variations in the instrumental profile or stability. However, the SVD/PCA based detrending step is effective at removing such correlated noise, often at the expense of needing a slightly higher value for the number of singular vectors ($k$) to be considered for the detrending step (see Section \ref{pca}). This holds true when the signature caused due to the variations responsible for correlated noise do not smear across multiple resolution elements. For USPs like 55\,Cnc\,e, the fast rate of change of the exoplanet’s radial velocity ensures that the SVD/PCA based detrending process can be pushed to relatively high $k$ (which we do indeed see in our analysis) without significant loss of the exoplanet signal. This provides an advantage for removing any correlated noise, even in a realistic scenario, without removing much of the exoplanet signal in the process. In addition, for a target star as bright as 55 Cnc A, the possibility of nonlinearity effects in the detector caused due to this high brightness can also induce correlated noise due to variations in quantum efficiency. This variation is harder to predict without a proper per-instrument characterisation. Hence, for all the reasons above, we have not included any explicit treatment for correlated noise in this study.

\subsection{Detecting an embedded exoplanet signal from simulated nights using \texttt{Upamana}} \label{pca}
Detecting the exoplanet signal we embedded in the flux cuboid generated above involves doing essentially the reverse of everything we did in Section \ref{nightsimul} and strip out every source of photon flux other than the time varying exoplanet signal. A way to do this is to try and remove all sources of flux variations that do not vary in time across wavelength since it's only the exoplanet signal which is being Doppler shifted across wavelength due to its orbital motion at a few km\,s$^{-1}$ throughout our observation. This same principle was used in \citet{dash2024constraints} to try and detect an H$_2$O signature in the atmosphere of a sub-Neptune GJ 3470\,b where a pipeline built using a SVD/PCA based framework was successful in detecting an injected signal at the level of the expected exoplanet signal strength in their case. In this work, we use the same pipeline (and name it as \texttt{Upamana}) to try and detect our embedded signal for 55 Cnc e, which is an even smaller exoplanet, from emission spectroscopic data. %In this work 
We only briefly outline the mechanism in which the pipeline works and point the reader to  \citet{dash2024constraints} for an in-depth description.

\subsubsection{Preprocessing through low flux masking} \label{prepro}
We first preprocess our flux cuboid by masking out the low S/N pixels and the pixels with the most strongly saturated features due to telluric absorption so that they don't influence the detrending process. This is done on each spectral order (i.e. on each $n_\mathrm{spectra}\times n_\mathrm{pixels}$ array) by setting a minimum flux threshold of 5 percent of the median of the highest 300 flux values in that spectral order, and then set the values of the pixels with photon fluxes below this threshold to 0. %This masks out the saturated telluric absorption features in each order. 
In subsequent sections, all analyses are assumed to be done order-wise unless stated explicitly.

\subsubsection{Standardisation and Detrending} \label{detrend}
\begin{enumerate}
    \setlength\itemsep{1em}
    \item Since it is the variations in flux rather than absolute flux values which ultimately matter for us, we first equalise the contributions from any large scale features that can contribute highly to flux variation. This is done by subtracting each pixel in a column of \textbf{A} (i.e. across time) through its mean, and then dividing all the values by the standard deviation of the corresponding pixel column before it was mean subtracted. This gives the standardised matrix $\textbf{A}_{\textbf{S}}$ where the flux data over each pixel (i.e. each wavelength) are given equal weights.
    \item The actual detrending procedure is then performed by using an SVD/PCA decomposition on $\textbf{A}_{\textbf{S}}$ followed by Multi-linear Regression (MLR) to capture the effect of scaling of the singular vectors due to prior standardisation. We assume that the first $k$ highest ranked singular vectors obtained by SVD capture all major sources of variation in the standardised matrix $\textbf{A}_{\textbf{S}}$ which do not shift in wavelength space, except for the time varying exoplanet signal. Since the value of $k$ is dataset specific, we provide more specific details in Section \ref{modeldet}. As in \citet{dash2024constraints}, the functional effect of the detrending procedure is assumed to be equivalent to multiplying the data matrix \textbf{A} by a (detrending) matrix $\beta$ resulting in the best fit matrix $\textbf{A'}$. $\textbf{A'}$ contains contributions from the first $k$ highest ranked singular vectors. At the end of this step, we are left with the matrix of residuals $\textbf{A}_\mathrm{\textbf{N}}$ with values clustered around 1.0, which is obtained as
    \begin{equation}
        \textbf{A}_\mathrm{\textbf{N},d} = \textbf{A} / \textbf{A'}.
    \end{equation} 
    The variations from 1.0 in $\textbf{A}_\mathrm{\textbf{N}}$ are due to the time varying exoplanet signal and noise. $\textbf{A'}$ and $\beta$ are saved to reproduce the effects of the analysis in this step on each forward model template used for cross-correlation as shown in the next step.
    \item As shown in \citet{dash2024constraints}, the detrending procedure outlined above leaves an impression on the embedded exoplanetary signal which needs to be accounted for during the cross-correlation analysis to provide accurate results. The effect of this is thus needs to be first reproduced on any model template used for cross-correlation. This is done by using the model template to first introduce an artificial exoplanet signal into $\textbf{A'}$ at the desired orbital solution corresponding to our exoplanet, which gives us \textbf{A}$_\mathrm{\textbf{m}}$. \textbf{A}$_\mathrm{\textbf{m}}$ thus has the time varying exoplanet signal as an additional singular vector apart from the $k$ highest ranked singular vectors used to reconstruct $\textbf{A'}$ in the previous step. Performing the same detrending procedure as done on \textbf{A} in the previous step should then leaves us with just the injected time varying model signal with the impression left on it by the detrending procedure. In matrix terms, we include the impression left by the detrending procedure by matrix multiplicating \textbf{A}$_\mathrm{\textbf{m}}$ by $\beta$ (as we did previously for the raw data matrix) to give the best fit matrix \textbf{A'}$_\mathrm{\textbf{m}}$. As in the previous step, \textbf{A'}$_\mathrm{\textbf{N},m}$ is the new matrix of residuals which now has variations from 1.0 due to the injected time varying exoplanet model only, alongside the impression produced on the exoplanet model template due to the detrending procedure. We employ the same masks calculated in Section \ref{prepro} and the resultant masked matrix is denoted as \textbf{A}$_\mathrm{\textbf{N},m}$.
\end{enumerate}

\subsubsection{Cross-correlation analysis through the CCF-to-$\log(L)$ framework} \label{ccf}
To calculate the similarity between \textbf{A}$_\mathrm{\textbf{N},d}$ and \textbf{A}$_\mathrm{\textbf{N},m}$ and concurrently provide a Bayesian estimator that can be used for parameter selection, we employ the CCF-to-likelihood framework devised in \citet{brogi2019retrieving} to calculate likelihood ($L$) values through the equation:
\begin{equation}
    \log(L) = -\frac{N}{2}\log[s_{f}^{2} - 2SR(s) + S^{2}s_{g}^{2}],
\end{equation}
$S$ is the scale factor by which the model is multiplied to seek a best fit and is a measure of how well the expected nominal model actually fits the data. For the best case scenario, when the model can correctly represent the data, as shown in in \citet{brogi2019retrieving}, its best fit value should be close to 1. In any retrievals, we set this as a free parameter (as $\log(S)$) to see how well our injected and retrieved models agree, which tells us about any biases inherent to the detrending procedure itself. In the Appendix A1, we perform a 3-parameter retrieval on one particular injected model case of $\Delta$IW-6 in Night 1 (see Sections \ref{specfug} and \ref{simnights} for details) where we are confident about detecting a signal, across 4 values of $k$ to check if $S$ is close to 1 in each case. We indeed find it to be the case. Henceforth throughout this study we simply set $S=1$. $s_{f}^{2}$ (data variance), $s_{g}^{2}$ (model variance) and $R(s)$ (cross-covariance) are defined as
\begin{eqnarray}
    s_{f}^{2} & = & \frac{1}{N}\sum_{n}f^{2}(n), \nonumber \\
    s_{g}^{2} & = & \frac{1}{N}\sum_{n}g^{2}(n), \nonumber \\
    R(s)      & = & \frac{1}{N}\sum_{n}f(n)g(n-s). 
\end{eqnarray}
$f(n)$ and $g(n)$ are the mean subtracted values of each row from each order of \textbf{A$_\mathrm{\textbf{N},d}$}, and the corresponding row from each order from \textbf{A$_\mathrm{\textbf{N},m}$}. $s$ denotes a wavelength shift, obtained by Doppler-shifting the model template for each tested value of exoplanet orbital velocity which is dependant on the particular exoplanet orbital solution used for injection in the reprocessing step in Section \ref{detrend}. 
\\
\\
The exoplanet orbital solution ($v_\mathrm{P}(t)$) for model injection is characterised by a $v_\mathrm{sys}$-$K_\mathrm{P}$ pair, as can be seen in Equation 10, where the other values are either known from literature ($e, \omega$) or calculated as part of synthesising the nights of observations ($v_\mathrm{bary}(t), f(t)$). Hence, detecting an embedded exoplanet signal means isolating the correct $v_\mathrm{sys}$-$K_\mathrm{P}$ pair used to inject the exoplanet model from a grid of $v_\mathrm{sys}$-$K_\mathrm{P}$ pairs. For this purpose, we choose $v_\mathrm{rest}$ (i.e. $v_\mathrm{P}-v_\mathrm{sys}$) from a grid of values between $-20$ and 20 km\,s$^{-1}$, equally spaced by 1\ km\,s$^{-1}$. $K_\mathrm{P}$ is chosen from a grid of values between 200 and 280 km\,s$^{-1}$ in intervals of 1\ km\,s$^{-1}$. In the SNR based metric widely used in HRCCS studies in literature, the spacing in v$_\mathrm{rest}$ generally has to be proportional to the Half-width Half Maximum (HWHM) of the instrument profile (e.g. 3 km$^{-1}$ at R=50000 for GIANO) according to the Nyquist sampling frequency, and oversampling can cause biases. However, for the likelihood based analysis we do here where we calculate significances from the likelihoods directly, we intentionally choose to slightly oversample by considering a spacing of 1 kms$^{-1}$ to maximise sensitivity and avoid the case of the peak of likelihoods falling between two grid values. For SNR analysis, the considered range of $v_\mathrm{rest}$ has to be large enough to get an unbiased measure of the noise needed to calculate the total SNR matrix from the CCF matrix. However, for our likelihood based analysis, this range only needs to be large enough to include any possible orbital configurations for the planet, and any possible combinations of systemic and barycentric velocity, since there is no explicit noise calculation. The only thing that needs to be considered is if the peak of the likelihood in $v_\mathrm{rest}-K_\mathrm{P}$ space corresponding to the detected model easily stands out compared to any other features in the considered range. 
\\
\\
With the grids now chosen, for each $v_\mathrm{rest}-K_\mathrm{P}$ pair, we first calculate a $\log(L)$ value by using each row for each order in \textbf{A$_\mathrm{\textbf{N},d}$} and the corresponding row in the same order in \textbf{A$_\mathrm{\textbf{N},m}$}. Then we sum across all rows in the order, and then sum across orders to calculate a single resultant $\log(L)$ value. This then leaves us with a matrix of $\log(L)$ values corresponding to exoplanet orbital solutions for each $v_\mathrm{rest}$-$K_\mathrm{P}$ pair in the grid. The highest value of $\log(L)$ ideally falls at the expected exoplanet orbital solution in the case of a detection. Then we utilise the same method employed in \citet{dash2024constraints} to calculate the confidence interval maps via a likelihood ratio test followed by application of Wilks' theorem \citep{wilks1938large} for 2 degrees of freedom, corresponding to a $v_\mathrm{rest}$ and $K_\mathrm{P}$ pair. A signal confidently detected appears as a series of tight, concentric contours around a certain $v_\mathrm{rest}$ and $K_\mathrm{P}$ pair, and the rest of the parameter space disfavoured by more than 4-5$\sigma$.
\\
\\
We extend the same approach to calculate $\log(L)$ matrices and confidence interval matrices for a grid of model templates generated using differing initial oxygen fugacities. Fugacities add another degree of freedom while calculating the confidence interval plots. The only difference in interpretation of such maps compared to detection maps is a rejection of models at the $n$-$\sigma$ confidence level, rather than rejection between orbital solutions.

%%%%%%%% MODELLING OF EXOPLANET SPECTRA WITH CHANGE IN OXYGEN FUGACITY %%%%%%%%%%%%%

\begin{figure}
    \centering
    \includegraphics[width = \columnwidth]{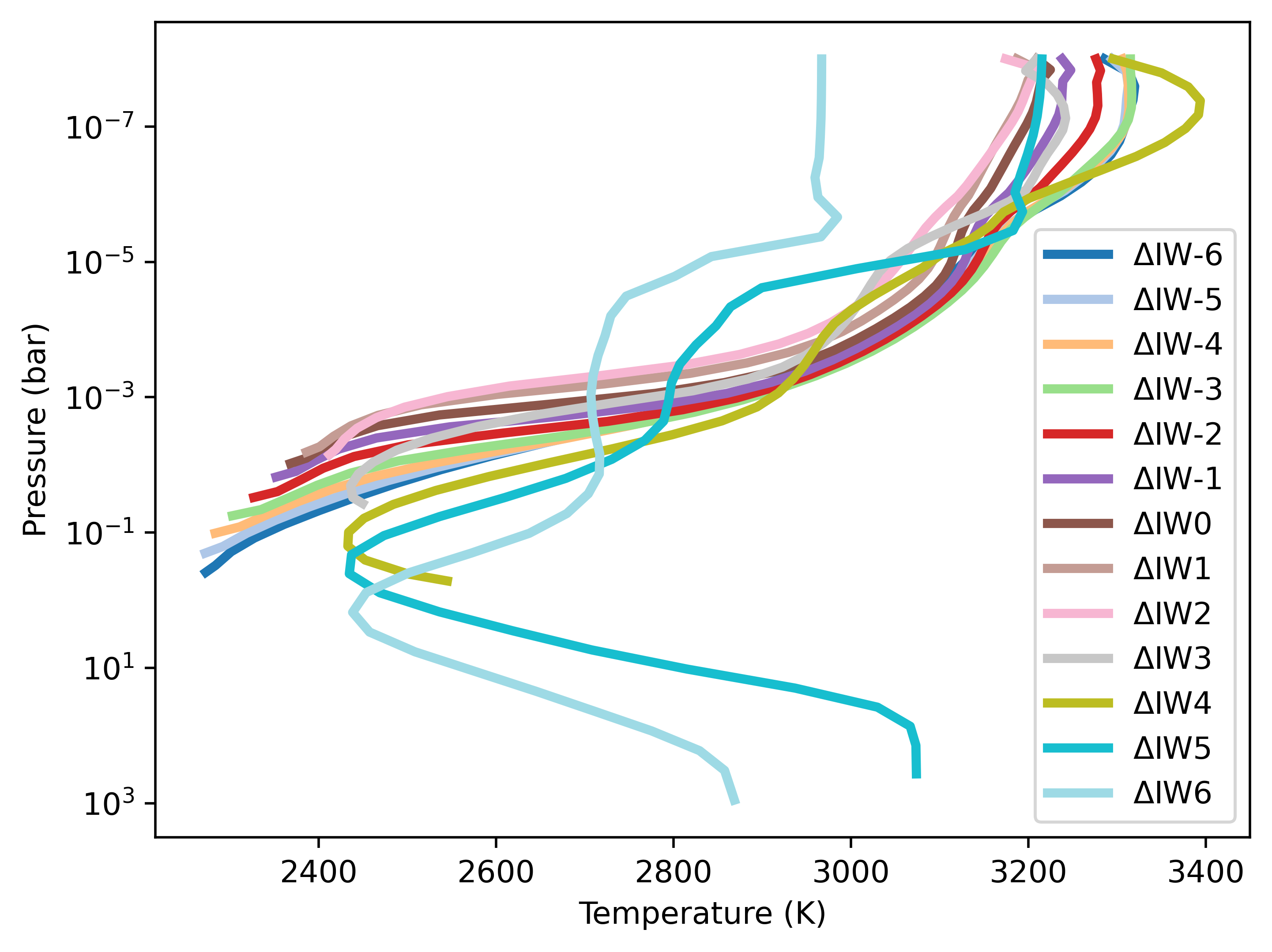}
    \caption{P-T profiles for the models used in this study as output from HELIOS.}
    \label{ptprof}
\end{figure}

\subsection{Modelling of exoplanet spectra for a grid of oxygen fugacities} \label{specfug}

\begin{figure*}
    \centering
    \includegraphics[width = 2\columnwidth]{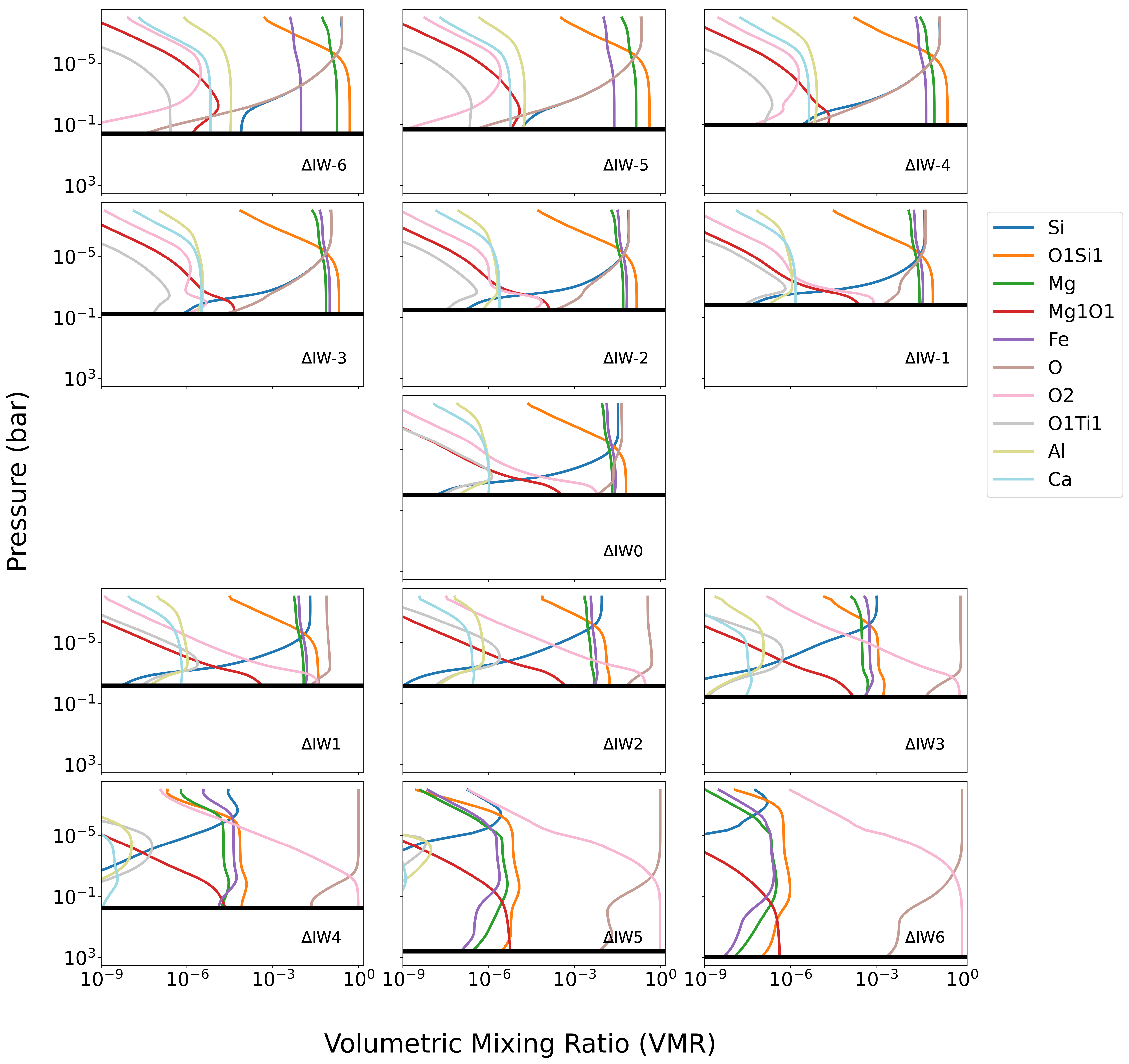}
    \caption{Abundances for the selected species to be used as inputs for GENESIS in Section \ref{genesis}. The black horizontal line in each panel represents the pressure at the surface of the exoplanet. For an explanation of the variation in thickness of atmospheres in the strongly reducing and strongly oxidising regimes, please see Section \ref{specfug}.}
    \label{abundances}
\end{figure*}

\begin{figure*}
    \centering
    \includegraphics[width = 2\columnwidth]{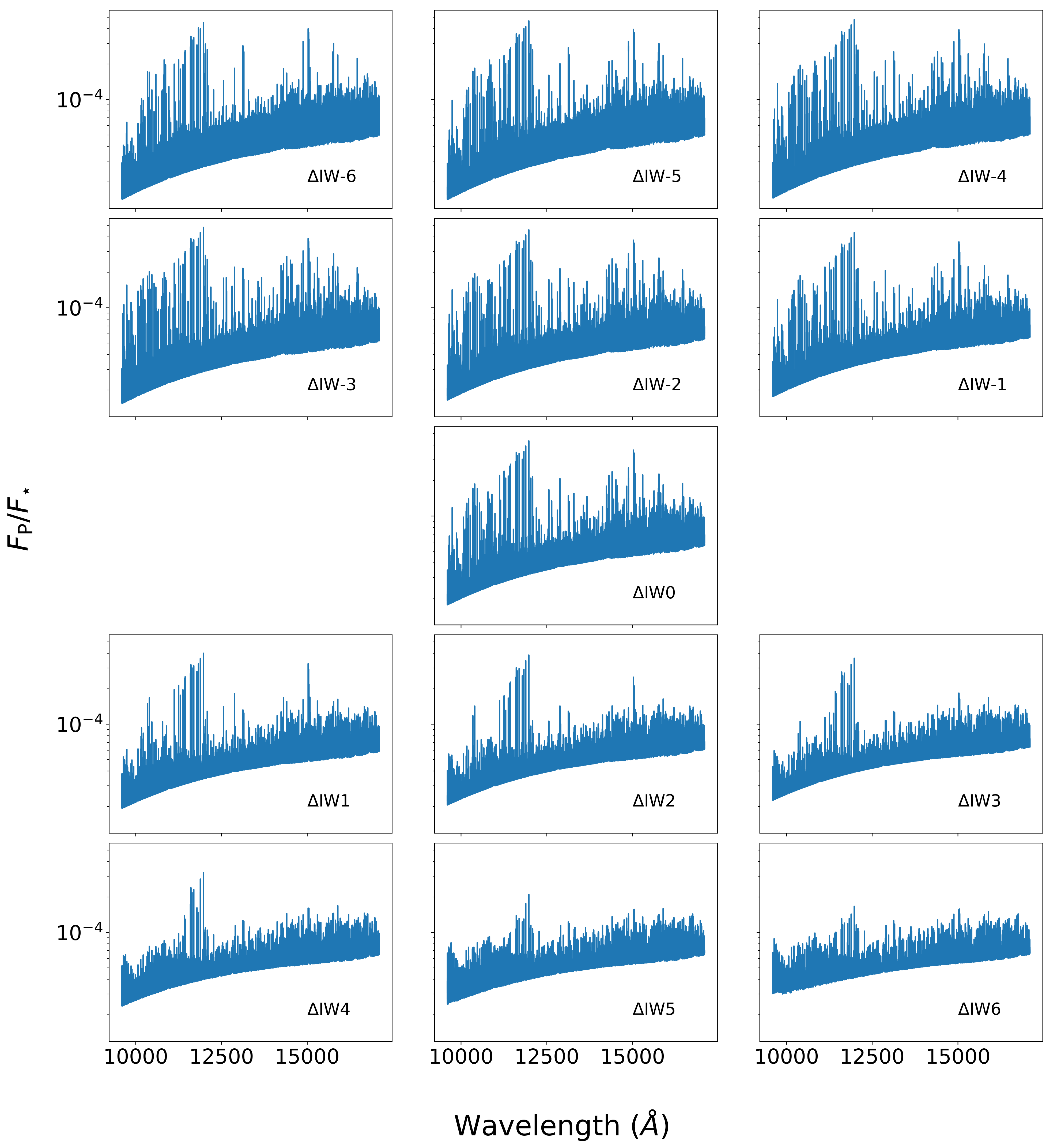}
    \caption{Output template emission spectra from GENESIS for all models used in this study. This has been restricted to the wavelength range covered by the infrared arm of CARMENES, since we use \texttt{Ratri} to only simulate CARMENES emission observations of 55\,Cnc\,e in this study.}
    \label{emtemplate}
\end{figure*}

%A substantial magma ocean, sustained by the extreme irradiation temperature of the host star, is expected to rapidly achieve chemical equilibration with the atmosphere. Under these conditions, the chemical potential $\mu$ of any species must be the same in the vapour as in the silicate, including oxygen. At a given pressure and temperature, the melt's \fOtwo is set by the relative activities of species of a redox pair, e.g. FeO and FeO$_{1.5}$ \citep{KressCarmichael1991, BerryONeill2021}. Because the mass of melt is assumed to be infinitely larger than that contained in the atmosphere, the \fOtwo imposed by the melt on the atmosphere fixes the identities and partial pressures of its constituent gas species. Because these species have different opacities, the resulting P-T structure of the atmosphere is sensitive to its composition  \citep{seidler2024impact}. 
%\\
%\\
We use the model of \citet{seidler2024impact} to derive the mixing ratios and pressure-temperature (P-T) structure of ``dry" mineral atmospheres, using the parameters for 55 Cancri e (Table \ref{tab:system_par}) and the stellar spectra for 55 Cancri A from \citet{zilinskas2022observability}\footnote{\url{https://github.com/zmantas/LavaPlanets}}. The melt is assumed to be of Bulk Silicate Earth (BSE) composition due to the unknown mantle compositions of (rocky) exoplanets and the redox chemistry of the melt is modelled using \texttt{MAGMA} as modified by \citet{seidler2024impact}. The speciation of the secondary mineral atmosphere in equilibrium with the melt as a function of pressure and temperature away from the interface is computed using \texttt{FastChem} \citep{stock2018fastchem}. For this study, we assume the case of a tidally locked planet with no heat redistribution, which corresponds to a dilution factor ($f$) $=$ $2/3$ (as in \citealt{seidler2024impact}), but also show the results for $f=1$ case in Appendix B. This sets the irradiation temperature at the melt surface, that is, at the bottom of the atmosphere (BOA) to be $\sim$ 2500K. The observations mentioned in Section \ref{55cncepuzzle} show that 55\,Cnc\,e exhibits large variations in the measurements of its brightness temperature and the irradiation temperature set here is well within this range. Using the \texttt{FastChem} outputs and the irradiation temperature, the radiative transfer code \texttt{HELIOS} \citep{malik2017helios} self-consistently solves for the atmospheric P-T profile by iterating the equilibrium speciation for the new P-T until they converge for a given irradiation temperature. As shown by \citet{seidler2024impact}, the atmospheric abundances and P-T-structure are far more sensitive to changes in \fOtwo than in composition when considering likely ranges in bulk mantle chemistry. 
\\
\\
Variations in atmospheric pressure-temperature structure over a grid of \fOtwo ranging from $\Delta$IW of -6 to +6 is shown in Figure \ref{ptprof}. The P-T profiles can vary from being thermally inverted over the entire atmosphere in the reducing cases ($\Delta$IW $< 4$), to the strength of thermal inversion being gradually reduced as the atmosphere becomes more oxidising ($4\leq\Delta$IW $\leq6$). As discussed by \citet{seidler2024impact}, the maximum pressure at the BOA also varies, with the thinnest atmospheres (10$^{-1}$ bar) occurring at \fOtwo near the IW buffer, while thickest atmospheres being the most reduced ($\Delta$IW-5, -6) and the most oxidised ones ($\Delta$IW+5, +6). The most oxidised atmospheres at $\Delta$IW+5, +6 are $\sim$3 orders of magnitudes thicker than the most reduced cases ($\Delta$IW-5, -6) due to - i) a strong greenhouse effect in O$_2$-rich atmospheres that leads to higher surface temperatures at a given irradiation temperature leading to higher $p$O$_2$ in equilibrium with the melt at constant $\Delta$IW, and ii) the positive, linear dependence of total pressure on $f$O$_2$ above $\Delta$IW$\sim$+2 (i.e. $P \propto fO_2$, as opposed to $P \propto (fO_2)^{-1/2}$ below $\Delta$IW$\sim$+2 which reflects the stoichiometry of the predominant vaporisation reactions; SiO$_2$(liquid) = SiO(gas) + 1/2O$_2$ and MgO(liquid) = Mg(gas) + 1/2O$_2$). The converged chemical abundance structure as the final result of self-consistent iterations between the melt, \texttt{FastChem} and \texttt{HELIOS} until P-T profile convergence is shown in Figure \ref{abundances}.
\\
\\
All reduced atmospheres have SiO, Mg and Fe forming the bulk of the lower atmosphere (P $>$ 10$^{-4}$ bar) with O and Si becoming dominant at the expense of SiO at P $<$ 10$^{-4}$ bar. The presence of metallic oxides like SiO, MgO, TiO and O (with SiO playing the dominant role) contributes to the strong thermal inversion as seen in their corresponding P-T profiles. Oxidised cases are typified by O$_2$ in the lower atmosphere (P $>$ 10$^{-2}$ bar) before it thermally dissociates to 2O below these pressures, with the other optically thick species (SiO, TiO, MgO) having lower mixing ratios as the \fOtwo increases. As stated in \citet{seidler2024impact}, $p$SiO decreases relative to $p$MgO as \fOtwo increases, making MgO the dominant molecule contributing to the atmospheric opacity. MgO absorbs strongly and nearly uniformly at $\geq$0.4$\mu$m, which makes these atmospheres optically thick throughout much of the infrared, but a decrease in mixing ratio of $p$SiO as $f$O$_2$ increases means that the atmospheres are no longer optically thick at $\leq$0.3$\mu$m \citep[below which SiO is a strong absorber, see][their Fig. 1]{seidler2024impact} Hence, more UV stellar irradiation can directly traverse most of the upper, optically thin atmosphere and heat up the bottom of the atmosphere. But, since the atmosphere is optically thick in the infrared due to the presence of MgO, the received radiation cannot be re-radiated into space. This trapping of radiation leads to a strong greenhouse effect in the most oxidising cases. Hence, the combined predominance of MgO opacity relative to that of SiO, and the reduced mixing ratio of both species, leads to a milder upper atmosphere thermal inversion (Fig. \ref{ptprof}), and to substantially thicker atmospheres for the most oxidising regimes.

\subsection{Generation of exoplanet emission spectra} \label{genesis}
The emission spectra of 55 Cnc e were computed via the \texttt{GENESIS} radiative transfer code \citep{gandhi2017genesis}, modified to correctly model a magma ocean surface and a secondary atmosphere. To this end, the reference pressure was chosen as the pressure at the magma-atmosphere interface (or at BOA which is variable with oxygen fugacity) and the planet's reference radius was set to the literature value of 0.167 $R_\mathrm{Jup}$ \citep{bourrier201855}. Such radius is also used with the exoplanet mass of 7.99 $M_{\oplus}$ to calculate the planet's gravity. The top-layer atmospheric pressure was set to 10$^{-8}$ bar and the atmosphere was divided in 50 plane-parallel layers equally spaced in log-pressure.
\\
\\
We used \texttt{GENESIS} to accept an altitude-dependent mean molecular weight (MMW), read directly from the \texttt{FastChem} outputs presented in Section \ref{specfug}. In terms of opacity sources, we only include the most prominent species, namely O, O$_2$, TiO, MgO, SiO, Al, Ca, Fe, Mg, Si. This subset is determined by discarding all the species with volume mixing ratios below 10$^{-5}$ at all altitudes, visually determined to clearly separate the most abundant and the least abundant species for all simulated compositions. The opacity line lists at high-resolution for all monoatomic species (O, Al, Ca, Fe, Mg and Si) were obtained from \citet{1995allbookKurucz}. The opacities for all metallic oxides were obtained from the ExoMol database \citep{tennyson2016exomol, tennyson20242024}. Specifically MgO was from \citet{li2019mgoexomol}, SiO is from \citet{yurchenkosio2022exomol}, TiO is from \citet{mckemmish2019exomol} and O$_2$ is from \citet{chubb2021exomolop} which is constructed using data from the HITRAN line list \citep{gordon2017hitran2016}. Additionally, we neglected FeO for which no accurate high resolution line list is available. The spectra were computed at a constant resolving power of $R=300,000$. Limiting the number of included species has the additional advantage that models can be calculated on a machine with relatively low RAM (16 GB in our case).
We note that we do not have pressure broadening coefficients available for any of the calculated atmospheric mixtures. Therefore, we ran the radiative transfer calculations with H-He broadening coefficients (assuming a mixture of 85\% of H$_2$ and 15\% of He), which may result in slight inaccuracies in the width of the modelled line wings. Nevertheless, as the pressures for almost all simulated atmospheres are low, pressure broadening is unlikely to influence the emission spectra. This is because even for the most oxidising atmospheres which can have surface pressures of $\sim$10$^3$ bar, the very high opacities of metal oxide gases, namely, MgO and SiO, result in photospheric pressures that are equal to or lower than $\sim$0.1 bar \citep[see Fig. 6 of][]{seidler2024impact} and the previous assumption would still hold. 
\\
\\
In its standard version, \texttt{GENESIS} would compute a Planck black body function at the stellar effective temperature of 5200 K and a stellar radius of 0.943 $R_{\odot}$ to scale the emission spectrum in units of stellar spectrum, which is appropriate to simulate high resolution spectroscopy where all the spectral features are expressed relative to the stellar continuum as a consequence of the data analysis presented in Section \ref{detrend}. However, since in these simulations we use a PHOENIX spectrum to simulate the observing nights, we opt to normalise all the GENESIS outputs by the same stellar PHOENIX spectrum to remove any possible normalization biases.

\section{Results}\label{results}
\subsection{Simulation of synthetic nights of observations} \label{simnights}
The 55\,Cnc system is a northern target with RA: 08 52 35.81, and Dec: +28 19 50.95 \citep{gaia2020vizier}, which makes it difficult to observe using CRIRES+ for most of the year since it is in the Southern hemisphere. Even in the small windows when the airmass values make an observation possible, obtaining a large number of exposures covering the dayside of the exoplanet would be a difficult task. In addition, \citet{dash2024constraints} pointed out that the SVD/PCA based detrending procedure is best served when the total amount of exposures obtained is large. Since obtaining enough exposures to maintain fidelity of the detrending procedure is a motivating factor, we use one of the spectrographs situated in the Northern hemisphere for this study, and we pick CARMENES as the representative spectrograph because it has the highest resolution and it is representative of many other similar spectrographs mounted on 3-4m class telescopes in the Northern hemisphere.
\\
\begin{figure}
    \centering
    \includegraphics[width = \columnwidth]{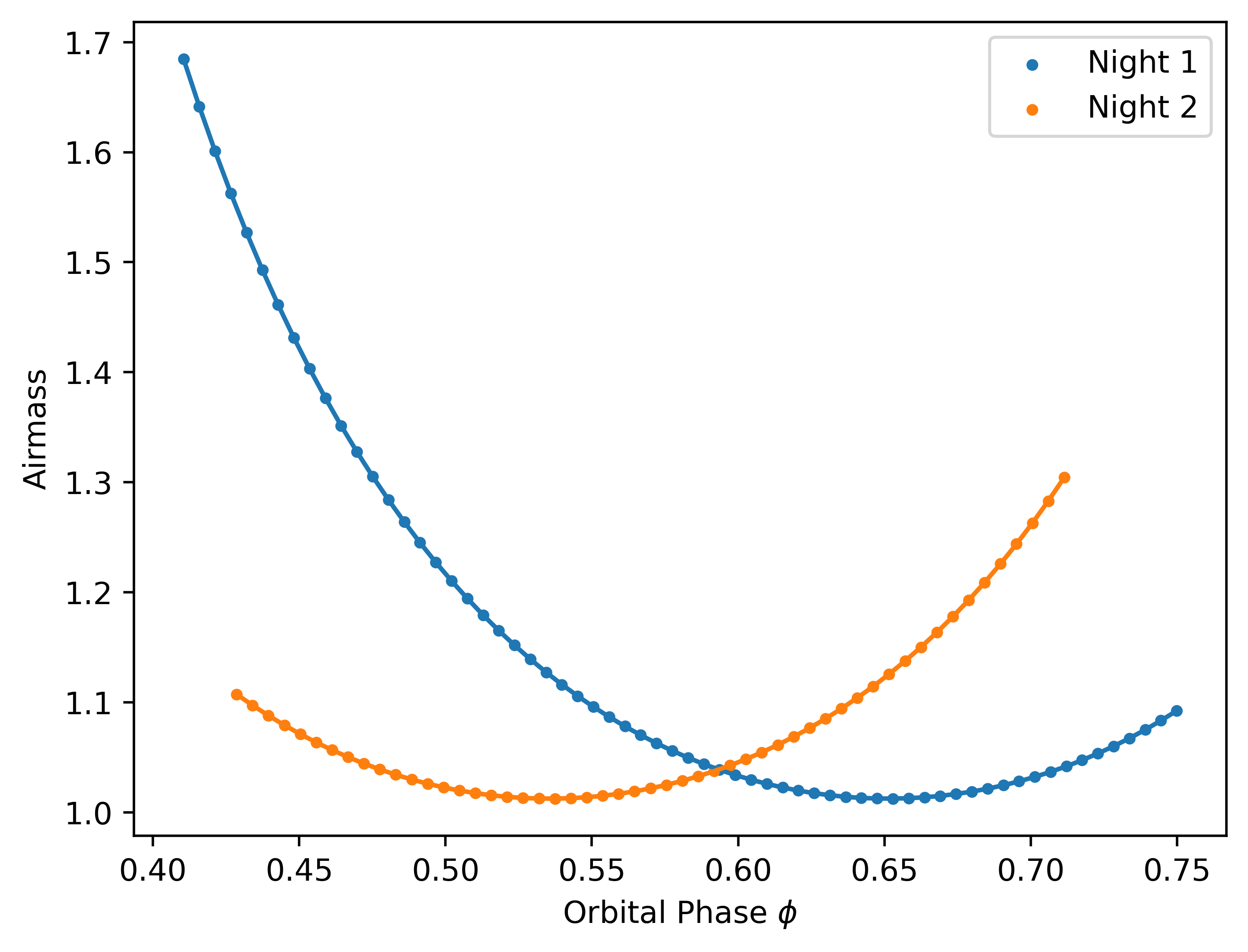}
    \caption{Airmass vs. exoplanet orbital phase ($\phi$) for both CARMENES nights simulated in this study.}
    \label{nights}
\end{figure}
\begin{figure}
    \centering
    \includegraphics[width = \columnwidth]{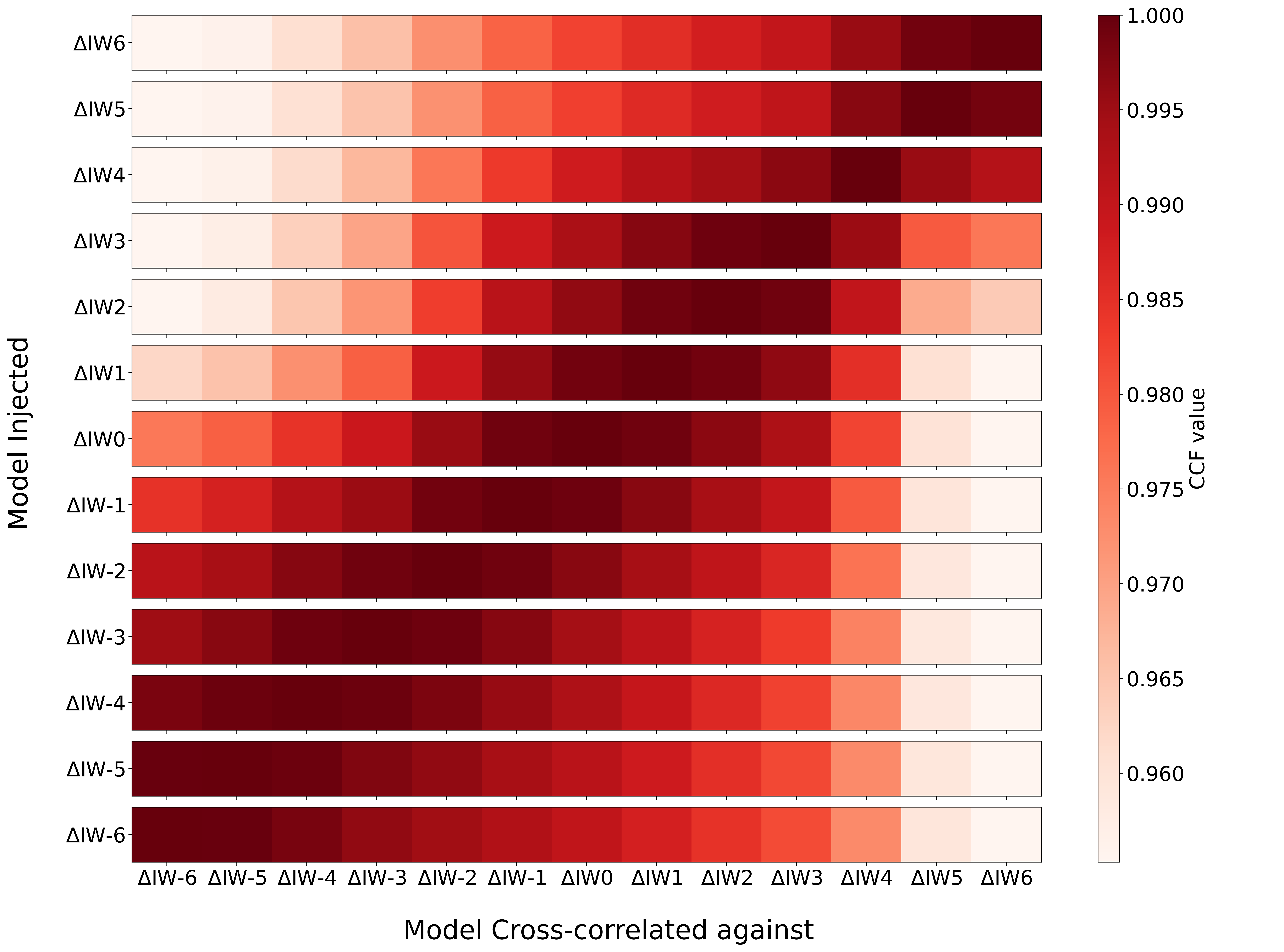}
    \caption{Similarity between the models depicted as the CCF values obtained between each set of models. A higher CCF value denotes a greater degree of similarity with a maximum value of 1.0. As expected, self-correlation produces the value of 1.0 and spans the diagonal across the lower left to the upper right corners. For details regarding the calculation of CCF values and classification of models, please see Section \ref{similarity}.}
    \label{ccf066}
\end{figure}
\\
We simulate two different nights of observations using \texttt{Ratri}: a 6 hour observation consisting of 64 exposures of 300s each starting from 22:00 UTC hours on January 2, 2024 and ending at 04:00 UTC hours on January 3, 2024 (hereafter denoted as Night 1) and, a 5 hour observation consisting of 53 exposures of 300s each starting from 20:30 UTC hours on March 3, 2024 and ending at 01:30 UTC hours on March 4, 2024 (hereafter denoted as Night 2). Both these nights were chosen because the range of airmass variations as shown in Figure \ref{nights} remains below 2.0 and the coverage of the dayside of the exoplanet is large enough while avoiding the points of quadrature at $\phi$ of 0.25 and 0.75, close to where the exoplanet signal will be stationary and hence almost entirely removed by the detrending procedure \citep[as seen for the case of Kelt-9\,b in][]{pino2022gaps}. As specified in Section \ref{telluric}, we designate a parameter of 'Good' for each night of observation; however to avoid any unrealistic discontinuous jumps in telluric absorption between 2.5 and 3.5 mm in the exposures, we use only one PPW value of 2.5 mm throughout each night. A more realistic treatment would be to find an empirical model to fit for measured water vapour variations at the site of each spectrograph to explicitly find the PPW values during each night of observation. While such an effect can be handled by our night simulator, we nevertheless do not consider that effect in this study. Hence, the variation in model telluric absorption during each of our nights would be due to the change in airmass value at each exposure, thus maintaining one major source of time variance of atmospheric conditions affecting each observation.
\\
\begin{figure*}
    \centering
    \includegraphics[width = 2\columnwidth]{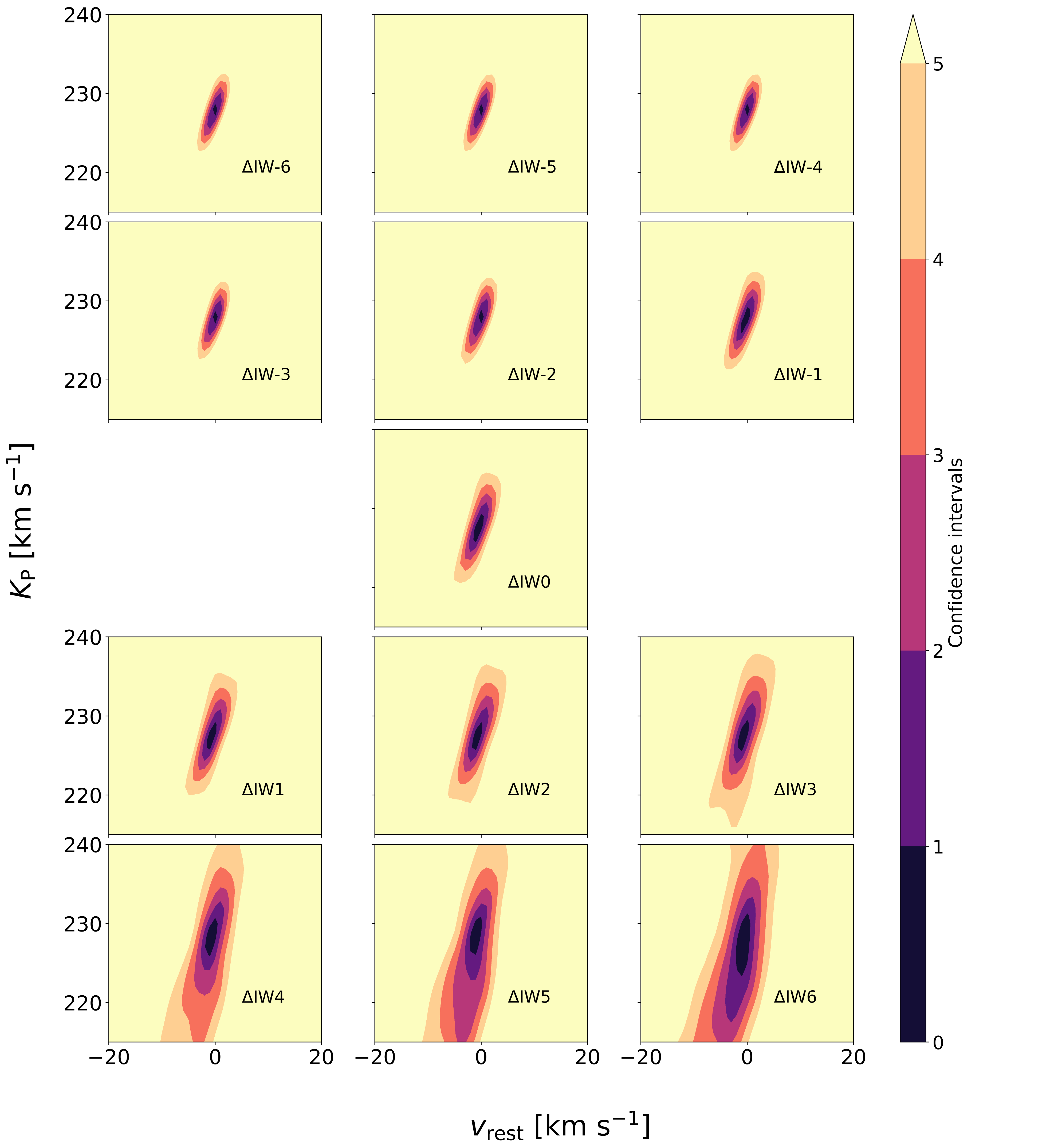}
    \caption{Recovery of the expected exoplanet signal for each fugacity regime on a $v_\mathrm{rest}$ - $K_\mathrm{P}$ grid, using two simulated nights of observations using CARMENES and $k=6$ for each night. For details regarding the nature of detection, please see Section \ref{modeldet}.}
    \label{conf066}
\end{figure*}
\\
Some portion of the dayside observation during each night would also be hindered by the exoplanet passing behind its host star (occultation) and hence not containing an exoplanet signal in the exposures. Assuming the occultation period to be the same as the transit period, and using values from Table \ref{tab:system_par}, we find that exposures taken between $0.45 < \phi < 0.55$ will be affected. This accounts for 16 exposures in Night 1 (leaving 48 exposures with an exoplanet signal) and 17 exposures in Night 2 (leaving 36 exposures with an exoplanet signal). While this effect is substantial, we neglect this effect of occultation in our study as we are primarily concerned with the total observation time (11 hours) looking at the exoplanet dayside rather than a perfectly realistic way of observing the exoplanet. Our modelled scenario of 2 nights would then represent an optimistic case of observation, and in reality we would need one more night of at least the same quality as Night 2 to get the same results as our simulation. Restricting our modelling to 2 nights also allows us to cut down on the computational time required to get our results.

\subsection{Similarities between models} \label{similarity}

We first examine the similarity between the emission spectra generated for our set of models, in order to quantify the degree to which they can be distinguished. In the perfectly noiseless case, equation 18 would be indeterminate as $f(n)=g(n)$ (where they now denote the respective mean subtracted values for each set of model) which will make the quantity inside brackets to be 0, which makes the likelihood calculation unsuitable for this purpose. That's why we simply compare the cross-correlation values (denoted as CCF) between each model obtained in Section \ref{specfug} and all models across \fOtwo (including itself) using the equation:
\begin{equation}
    \mathrm{CCF} = \frac{R(s)}{\sqrt{s^{2}_{f}s^{2}_{g}}},
\end{equation}
where the terms are the same as defined in Equation 19. A CCF value of 1.0 represents perfect similarity and should correspond to an auto-correlation of the model with itself, and values close to 1.0 represent a high degree of similarity. Figure \ref{ccf066} shows CCF values obtained by cross-correlating each model (y-axis) with the entire set (x-axis). As expected, a value of 1.0 spans the diagonal from the lower left to the upper right corner indicating the auto-correlation axis. In general, there is a high degree of similarity among the models with the minimum CCF value being $\sim$ 0.96. However, the plot also shows that the autocorrelation diagonal is bounded on each side by models which give CCF values $\geq$ 0.99 with each other, and hence are expected to be difficult to differentiate from each other. %is also sub-divided into three families of models that give CCF values $\geq$ 0.98 with each other, and hence are expected to be difficult to differentiate. The reducing models with \fOtwo between $\Delta$IW-6 and $\Delta$IW0 show a high degree of similarity with each other and form the first family. The mildly oxidising models with \fOtwo $\Delta$IW+1 to $\Delta$IW+4 form another such family, and the strongly oxidising models with \fOtwo $\Delta$IW+5 and $\Delta$IW+6 form the third family. 
Hence, even with our simulations we do not expect perfect differentiability for each model where the model selection trend will lie perfectly along the autocorrelation diagonal, but it will rather be a bounded detection of each model within the closely matched set of models that lie close to the autocorrelation diagonal. %We also expect the granularity of detection to be higher for strongly reduced fugacity regimes, compared to both mildly and weakly reduced and oxidised regimes and strongly oxidised regimes.

\subsection{Model detection} \label{modeldet}
The similarity between the models themselves, however, is not indicative as to whether the same models are detectable using the 2 synthetic nights. Equation 20 is invariant under any scaling factor ($S$) used, which means that the CCF is high as long as the relative line ratios (and their positions) are correct, even if two models differ by a very large overall $S$. The latter, however, changes the level of detectability of the signal. 
\\
\\
To showcase the detectability of all our models, we use 6 components (i.e. $k=6$) for each night to perform the SVD/PCA based detrending step of using \texttt{Upamana} as seen in Section \ref{detrend}, and then plot the confidence intervals calculated from a $\log(L)$ matrix constructed in $v_\mathrm{rest}$-$K_\mathrm{P}$ space as mentioned in Section \ref{ccf}. We note that the strongest model is detectable for a range of $k=4$ to $k=10$ (see Appendix A1 for an example). Our choice of $k=6$ here is an arbitrary selection for an illustrative purpose between these two values, where most of our models are indeed successfully detected (see below). We first calculate a $\log(L)$ matrix for each night and then subsequently add them together to get the combined likelihood across both nights. We also discard the spectral orders numbering 8-10 and 18-22 (i,e, order IDs starting from 0, not the actual order numbers\footnote{Order ID 0 corresponds to actual order 63 and ID 27 to actual order 36. Details of spectral orders with original order numbering can be found here: https://carmenes.caha.es/ext/instrument/DeltaLambdaNIR.txt}) in both nights before our analysis as they are marked by heavy telluric line saturation.
\\
\\
We show the obtained confidence interval plots from the combined $\log(L)$ matrix in Figure \ref{conf066}. While we calculate the $\log(L)$ values between 200 to 280 km\,s$^{-1}$ in intervals of 1 km\,s$^{-1}$ in $K_\mathrm{P}$ space, we narrow the window down to lie within  215 to 240 km\,s$^{-1}$ for these plots to show the detections properly. All models are prominently detected using our two synthetic nights. However, the nature of the detections vary, with the reducing and oxidising \fOtwo regimes with the strongest line core strengths ($\Delta$IW-6 to $\Delta$IW+3) producing detections across the 2 synthetic nights at comparable velocities in the $v_\mathrm{rest}$-$K_\mathrm{P}$ map. This perfect alignment allows for a very tightly constrained combined unimodal detection at the exact position of an expected exoplanet signal. The constraints on detections start loosening slightly as the model line core strength reduces ($\Delta$IW+4 to $\Delta$IW+6). The equivalent detection maps for the T$=$3000K case is shown in Appendix B (Figure B5). 

\subsection{The perfect detrending case} \label{optdetrend}
\begin{figure}
    \centering
    \includegraphics[width = \columnwidth]{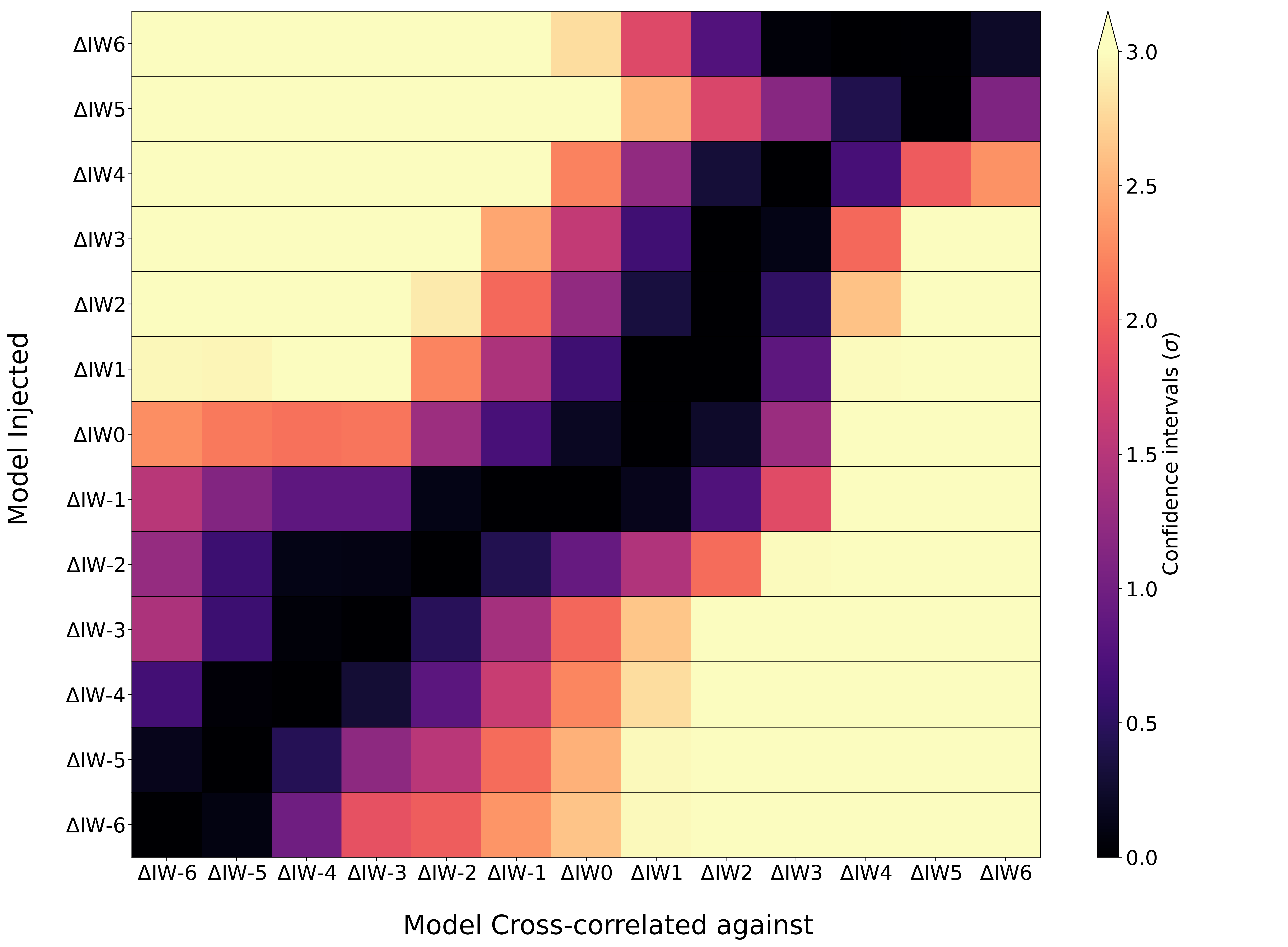}
    \caption{Grid based model Selection in the perfect detrending case. Injected models are selected within the 1$\sigma$ contours.}
    \label{optsel066}
\end{figure}
Now that we have illustrated that all models in our grid are detectable for $k=6$ specifically, but mentioned that the trend holds more generally from $k=4$ to at least until $k=10$, we perform a test to see what would be the best case scenario for their differentiability. In Section \ref{detrend}, we used a SVD/PCA + MLR based strategy to obtain the best fit matrix \textbf{A'} to detrend the data matrix \textbf{A}. In an optimistic scenario, we assume that we know \textbf{A'} exactly by retaining the noiseless flux cuboid before the model injection step for this purpose and label it as \textbf{A'}. \textbf{A}$_\mathrm{\textbf{N,d}}$ would then comprise only the exoplanet signal and the white noise as all the variations due to stellar and telluric contributions would be exactly divided out. Hence, this section is a case study to understand the best-case scenario of model differentiability. Since the detrending here would be exact and is just a simple division, there is no need to do an additional model reprocessing step and we instead do a direct CCF-to-$\log(L)$ analysis with the template model to calculate the $\log(L)$ matrix.
\\
\\
For each of the models, we now repeat the detection procedure for the two synthetic nights of observations as in Section \ref{modeldet} but with the perfect detrending procedure as illustrated above. We also limit the $v_\mathrm{rest}$ values to be between [-5, 5] km\,s$^{-1}$ with spacing of 1 km\,s$^{-1}$, and $K_\mathrm{P}$ values to be between [215, 240] km\,s$^{-1}$ with spacing of 1 km\,s$^{-1}$. This is done to minimise the computational time and to focus only in the velocity space where the exoplanet signal should arise as seen from Figure \ref{conf066}. After calculating the $\log(L)$ matrices on the $v_\mathrm{rest}$ and $K_\mathrm{P}$ grids defined above, for each model, we store the highest likelihood value in a grid corresponding to the position of its \fOtwo. For each injected fugacity case, we would have 13 $\log(L)$ values obtained by cross-correlation analysis with all models in our fugacity grid. The likelihood-to-confidence intervals approach mentioned in Section \ref{ccf} is then used to calculate the confidence interval contours for this grid-based model selection framework, but now with 3 degrees of freedom used in calculation. The extra degree of freedom here compared to that of Section \ref{modeldet} is due to the fact that \fOtwo is now an additional free parameter. Doing this for all models would then give us a 13$\times$13 grid as shown in Figure \ref{optsel066}. This particular method of grid based model selection is similar to the grid based analyses performed in \citet{lafarga2023hot} and \citet{dash2024constraints} to parametrise water abundance and cloud deck pressure levels, but with the difference that the models in the grid all have different P-T profiles computed by self-consistent calculations as explained in Section \ref{specfug}. 
\\
\begin{figure*}
    \centering
    \begin{tabular}{cc}
        (a) $k=4$ for each night & (b) $k=6$ for each night \\
        \includegraphics[width = \columnwidth]{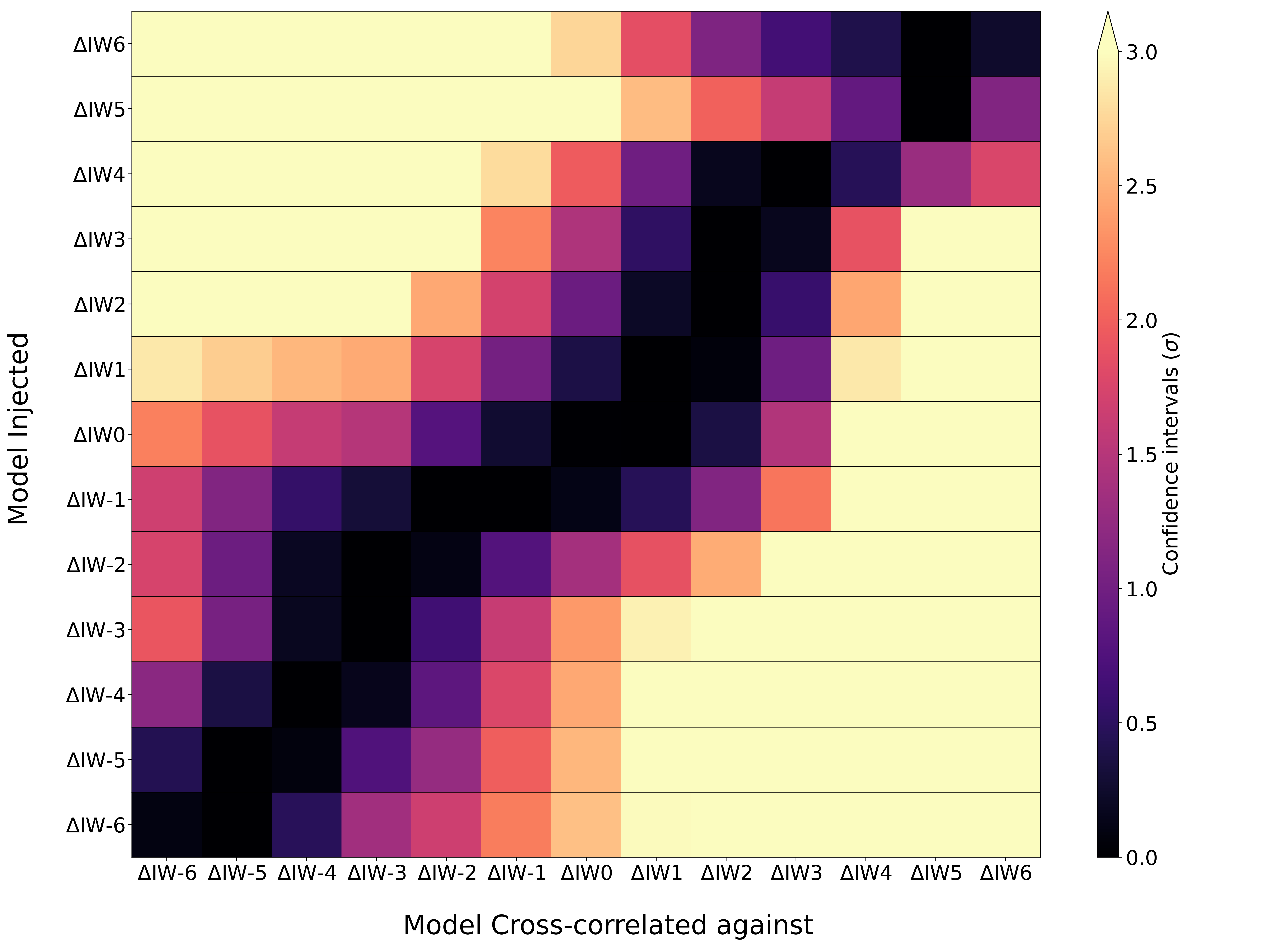}  & \includegraphics[width = \columnwidth]{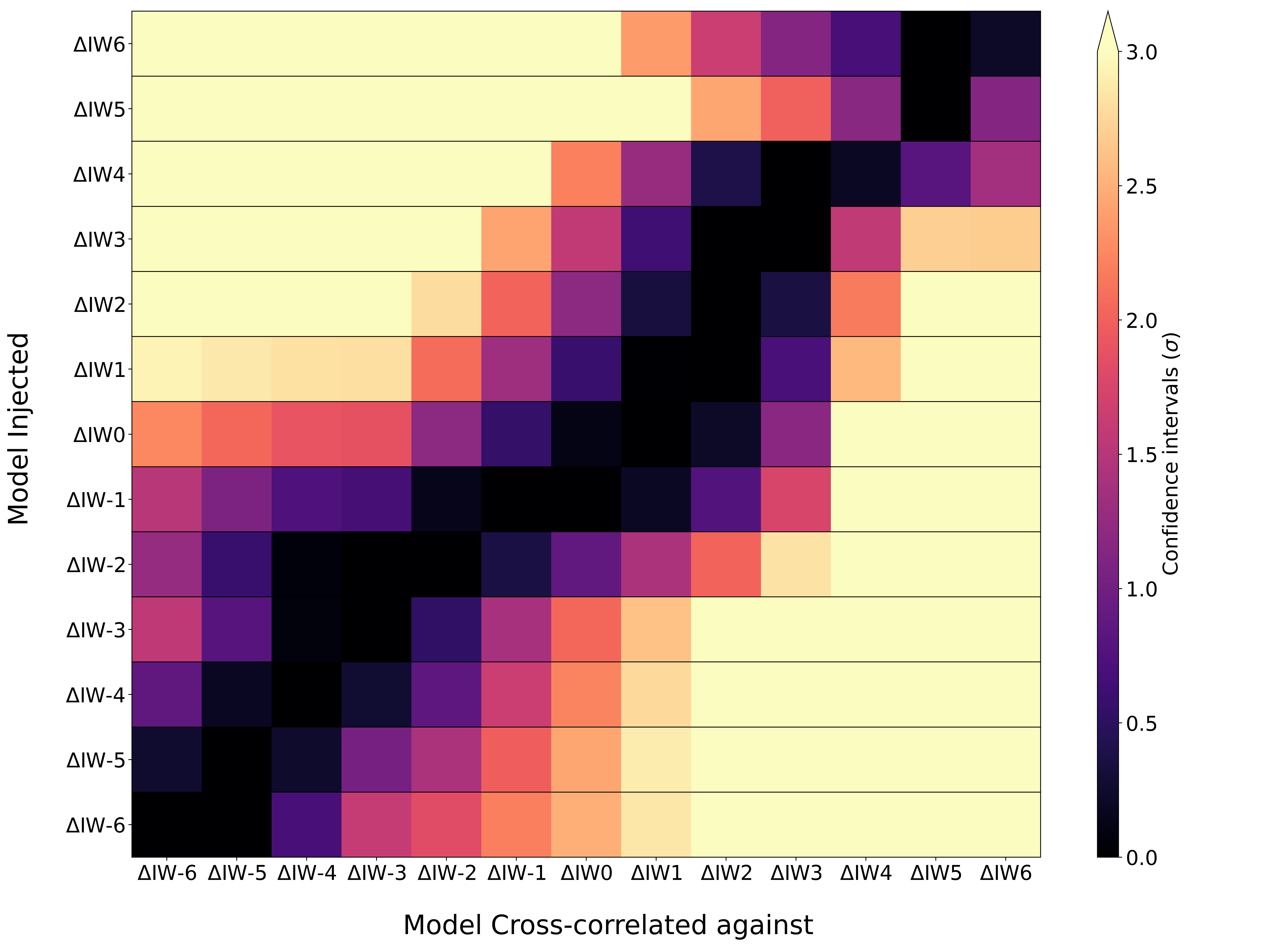} \\
        (c) $k=8$ for each night & (d) $k=10$ for each night \\
        \includegraphics[width = \columnwidth]{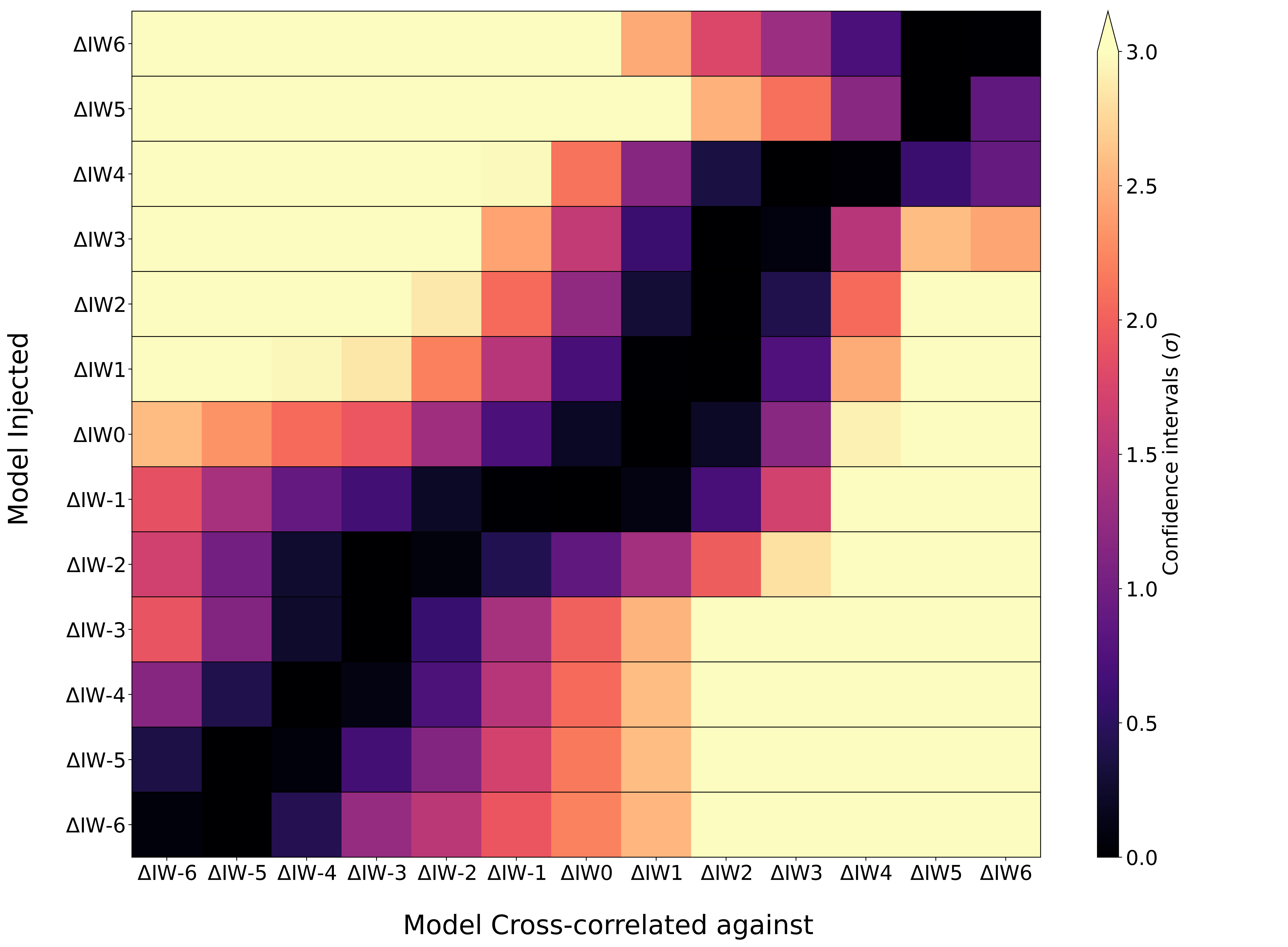} & \includegraphics[width = \columnwidth]{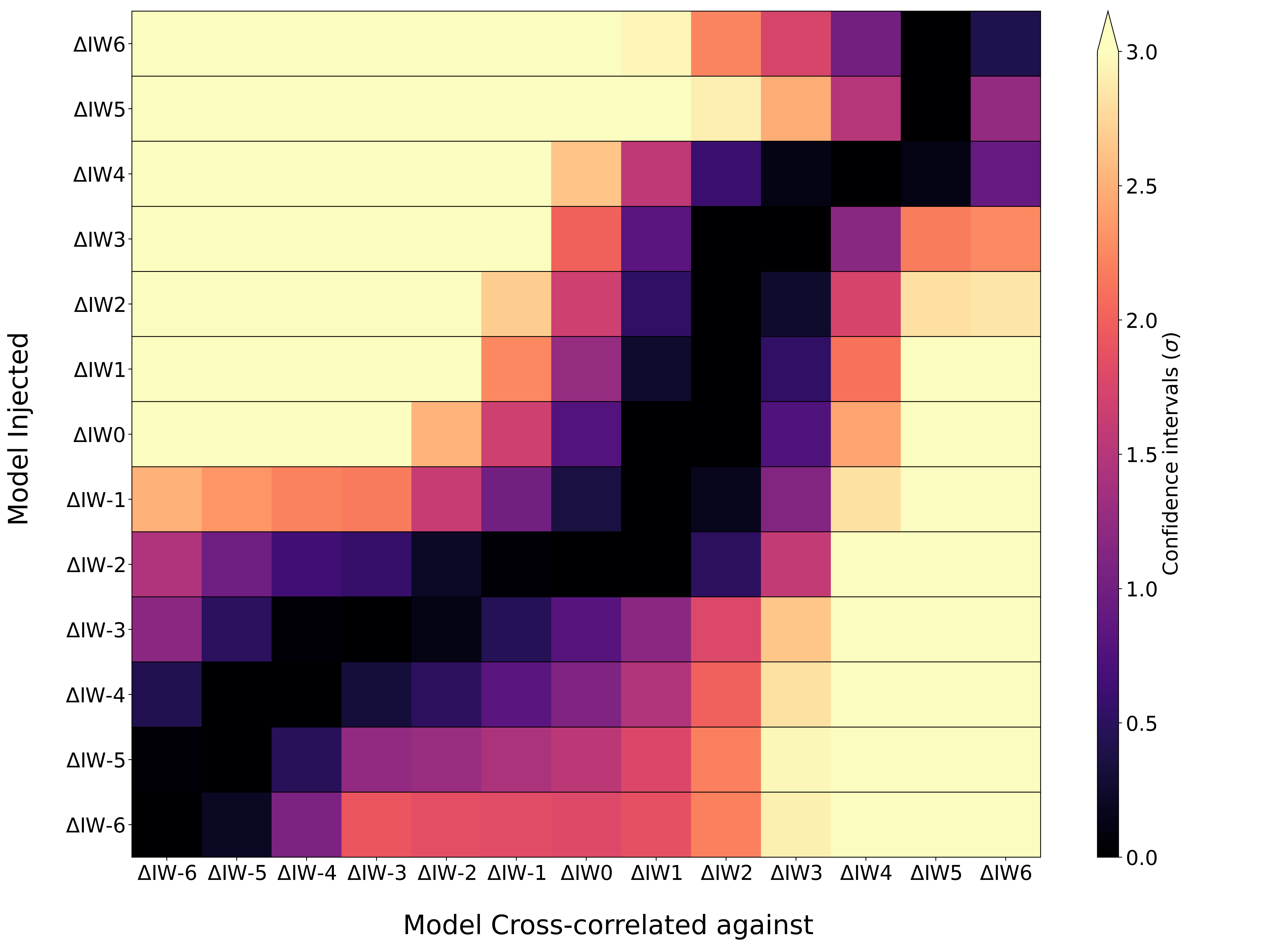} \\
    \end{tabular}
    \caption{Grid based model Selection in the SVD/PCA based detrending case for 4 values of $k$ used for detrending each night of observation. Injected models are always selected within the 1$\sigma$ contours in all cases.}
    \label{svdsel066}
\end{figure*}
\\
Figure \ref{optsel066} shows that the model being injected is the one retrieved within 1$\sigma$ contours for all cases. Models with close CCF values ($\geq 0.99$) are difficult to differentiate from each other and hence the model selection procedure sometimes can choose one closely matched model over the actual injected model. Nevertheless, as expected, the model selection is not perfectly along the auto-correlation diagonal but in a thicker set of models approximating it. Since all the reducing models ($\Delta$IW-6 to $\Delta$IW-1) are very similar to each other (CCFs $\geq 0.99$), the constraints on the diagonal selection are looser for such models. The constraints become tighter for the predicted IW and moderately oxidising model cases ($\Delta$IW0 to $\Delta$IW+4), and then loosen again for the highest oxidising case ($\Delta$IW+6). The highest oxidising models have the weakest line core strengths and from Figure \ref{ccf066} show a high degree of similarity to all oxidising models. A combination of these two factors might explain the longer tail of confidence in the cross-correlated models away from the injected $\Delta$IW+6 case. 

\subsection{SVD/PCA based detrending with reprocessing} \label{svddetrend}
Having now understood the best case scenario of model \fOtwo differentiation, we now repeat the same procedure as above but instead use the SVD/PCA + MLR based detrending procedure followed by model reprocessing using \texttt{Upamana} as outlined in Section \ref{detrend}. The 13$\times$13 plot corresponding to this model selection procedure is shown in all four panels of Figure \ref{svdsel066}, where each panel corresponds to removing flux variation contribution corresponding to the highest 4 (upper left), 6 (upper right), 8 (lower left) and 10 (lower right) components during the detrending procedure for each simulated night to showcase the effect of $k$ on the efficacy of the grid based model selection procedure. All four panels show model selection trends close to the one obtained using a perfect detrending procedure, and have the injected model \fOtwo being the one retrieved within 1$\sigma$ contours in all cases. While the $k=4$ to $k=8$ panels are overall very similar to the perfect detrending case in the span of the contour tails away from the auto-correlation diagonal, the $k=10$ panel showcases a trend with slightly worse constraints for the reducing fugacity regimes. Sometimes the selection of models corresponding to the highest oxidising regimes, especially for $\Delta$IW+6 is marginally better compared to the perfect detrending case. But it is important to understand that while comparing between the results from the perfect and SVD/PCA based detrending, the focus should be on the overall ability to distinguish between all models, rather than on the trend for each model - the latter being subject to marginal variations in significance due to the aleatory nature of statistical tests. With this perspective, all the model selection plots with SVD/PCA based detrending have overall slightly wider confidence intervals for reducing regimes even when they might perform marginally better for the highest oxidising regime. Nevertheless, this analysis showcases the efficacy of the blind SVD/PCA based detrending and model reprocessing procedures used in \texttt{Upamana} as the selection results using this pipeline are almost comparable to what is obtained from a perfect detrending scenario. Hence, our simulations have shown that 11 hours of CARMENES observations is not only sufficient to detect the diverse mineral atmospheres predicted for 55\,Cnc\,e, but is also enough to be able to differentiate between different oxygen fugacity regimes. 

\section{Discussion}\label{discussion}
\subsection{Constraining oxygen fugacity of rocky exoplanets with HRCCS}
\citet{seidler2024impact} had posited that the \fOtwo of 55\,Cnc\,e could be estimated by examining the relative intensities of SiO (at 9$\mu$m)-MgO (at 6$\mu$m) spectral feature contrast ratio in emission, since that ratio was found to be a monotonically decreasing function with increase in fugacity for a constant temperature and composition. 5 occultations observed in the mid-infrared (MIR) using JWST-MIRI could indicate the fugacity to a precision of $\pm$1 log unit. While there were no direct simulation based inferences, they also postulated that the visible region could also be amenable to determining the \fOtwo, as it would have spectral features arising from TiO, Mg and MgO. The mineral atmospheres \citet{seidler2024impact} simulated were featureless in the NIR window at the resolution pertinent for JWST.
\\
\\
In this study, we show that the HRCCS based analysis of 11 hours of synthetic observations in the NIR window (0.8-2.5 $\mu$m) using a representative current ground-based spectrograph in the Northern hemisphere (the infrared arm of CARMENES) of the dayside of 55\,Cnc\,e, permits identification of different kinds of mineral atmospheres and constraints to be placed on their \fOtwo (Section \ref{svddetrend}). While \citet{seidler2024impact} used intensity contrast in the spectral features of two specific molecules, here, our approach differs in that it exploits a forest of lines to quantify how similar or dissimilar these atmospheres are to each other. 
\\
\\
The range of oxygen fugacities thought to characterise the exoplanetary population are poorly constrained. Studies of polluted white dwarfs, which are based on the  abundance of O relative to other rock-forming elements (chiefly Fe, Mg and Si) are used to infer plausible \fOtwo ranges among rocky exoplanets \citep{doyle2019oxygen,harrison2021evidence}. This proxy, though instructive, does not provide an estimate of \fOtwo \textit{sensu-stricto}, as its computation relies upon the knowledge of pressure and temperature, rather than composition alone as deduced from polluted white dwarfs. For example, the Earth contains less O as a fraction of its bulk composition than does the Moon, yet, its upper mantle defines a higher \fOtwo \citep[$\Delta$IW+3.5, from][]{frostmccammon2008} than does that of the Moon \citep[$\Delta$IW-1, from][]{wadhwa2008}. Nonetheless, first order constraints on oxygen abundances indicate exoplanetary bulk compositions similar to those of the Earth or Mars \citep{doyle2019oxygen}, with inferred \fOtwo between $\sim \Delta$IW-2.5 and -0.5.
\\
\\
Were this range of \fOtwo to be representative of that expected in exoplanetary atmospheres, our HRCCS analysis illustrates that the spectral properties of such atmospheres would be barely resolvable from one another. However, other factors may act to broaden the spread of the small \fOtwo range inferred from measurements of polluted white dwarfs. In addition to the bulk compositions of planetary building blocks, \fOtwo at the \textit{surface} is known to be a function of the pressure of core-mantle equilibration pressure \citep{armstrong2019,deng2020}, whereby pressures exceeding $\sim$10 GPa will produce significant quantities of ferric iron (Fe$^{3+}$) in the silicate mantle of the planet. That of the Earth, whose core is thought to have formed at average pressures near 40--50 GPa \citep{siebert2013} contains 3.5~\% Fe$^{3+}$ \citep{sossi2020}, whereas  Fe$^{3+}$ is absent in that of the Moon.  Should the mantle physically homogenise (e.g., via convection), then it can be approximated as isochemical. The non-negligible quantity of Fe$^{3+}$ \citep[up to $\sim$20-30~\% at 60--70 GPa;][]{zhang2024ferric} would result in \fOtwo at the surface in excess of IW+3.5 (by analogy with that of the Earth's upper mantle). Indeed, the prevailing mineralogy of the mantle itself, even at constant Fe$^{3+}$/Fe$^{2+}$ ratio can result in a further spread in \fOtwo \citep{guimond2023}. On the other extreme, evidence for bodies that inherit the \fOtwo of the solar nebula is provided by Mercury, whose interior is thought to define an \fOtwo of $\Delta$IW-5.4$\pm$0.4 \citep{namur2016}. Hence, the \fOtwo defined by atmospheres that are produced in chemical equilibrium with the upper mantles of such planets would span $>$ 8 orders of magnitude, and would thus be readily detectable using our approach. 
\\
\\
At present, ground-based observations in emission have failed to detect any gaseous species that would testify to the presence of a magma ocean on 55\,Cnc\,e (see Section \ref{55cncehighres}), but as also explained in that section the template spectra used for cross-correlation in such studies were different compared to the ones derived by our framework where the coupling of the surface magma melt to the atmosphere is explicately considered. Hence, we hypothesise that with our new HRCCS framework, predictions as to the expected distribution of gaseous species will aid in the detection of magma ocean-derived signals. To place even tighter constraints, more nights of observation would be one possible avenue and a combination of using both high- and low-resolution data could be another as low-resolution observations will additionally provide information about the continuum of the emission spectrum, something which is lost during HRCCS detrending. If there is indeed coupling between the atmosphere and the magma melt in the way assumed in this study, the similarity between template spectra with very different oxidation states ensures that at least some level of detection should be achievable with any spectral template with future ground-based high-resolution studies. However, as mentioned in the penultimate point in Section \ref{limitations}, volatile-rich atmospheres are also one possible case of atmospheres on this exoplanet based on JWST observations. Such atmospheres would be cooler and have much fainter line cores, requiring facilities significantly larger (e.g. ELT-like) than currently available, to be detectable. So a non-detection of any template spectra generated with our framework could point to the presence of a volatile-rich atmospheric regime.

\subsection{Limitations of this study} \label{limitations}
\begin{itemize}
    \item For constructing the synthetic nights used in this study, we have used only a single PWV value of 2.5 mm for each night but realistically it will also vary across the observation time period. While \texttt{Ratri} can handle time varying PWV values, the exact way they will vary is something which will depend on each location considered \citep{foster2006precipitable}, which is why we did not assume an ad-hoc way to incorporate PWV variation in this study. A more realistic variation of PWV is very likely to be taken care of by using a slightly higher value of $k$, because SVD/PCA based detrending procedures are known to very efficient at removing telluric contamination \citep[see ][for an example]{meech2022applications}. We have shown that our HRCCS analysis provides robust model selection results across a wide range of $k$ (starting from 4 to at least 10). We thus expect the results to stay true even for a more realistic sequence of PWV variation in the Earth's atmosphere.
    \item A limitation in the SNR calculation using \texttt{Ratri} is the restriction to a photon noise dominated regime. Since 55\,Cnc\,A is an extremely bright star, simulating only the photon noise and neglecting dark current and readout noise make sense here. Addition of such sources is thus unlikely to change the results we obtain in this study, and should hence stay true for actual observations. However, both these neglected contributions can be easily added as additional white noise terms in future work on fainter targets, where neglecting them can results in overestimation of the significance of detections and selections.
    \item A factor that concerns both the generation of emission models (through the F$_\mathrm{P}$/F$_{\star}$ normalisation) used in this study and construction of the synthetic observations using \texttt{Ratri} is the stellar spectrum used. We have used a 1D PHOENIX model for the stellar spectrum in this study which assumes a spherically symmetric and homogenous photosphere, but is not entirely realistic. In actual observations, 3D effects like flaring or long term variability that can change that level of normalisation, will affect the efficiency of the detrending procedure. \citet{folsom2020circumstellar} found that 55\,Cnc\,e could be orbiting inside the sub-Alfvenic region of the stellar wind, and star-exoplanet magnetic interactions might hence lead to the origin of stellar activity \citep{valdes2023investigating}. However, \citet{valdes2023investigating} ruled out this possibility, and also found that other star-exoplanet interactions like tidal interactions were also unlikely to be a source of enhanced stellar activity. So, based on current evidence, we don't expect the level of normalisation to vary a lot during the time period of actual observations and influence the results obtained in this study.
    \item As discussed in Section \ref{55cncepuzzle}, mineral atmospheres are consistent for only some of the observed emission spectra (including some JWST spectra) for 55\,Cnc\,e. Results from other JWST spectra show that CO/CO$_2$-rich atmospheres appear to be required to fit the prominent absorption at $\sim$4.5 $\mu$m. Such atmospheres would have much fainter line cores in emission, and hence detecting them would require the next generation of ground-based facilities. Here, we are concerned only with mineral atmospheres, and the case of volatile-bearing (H, C, N, S) atmospheres are neglected, but nevertheless represent a tantalising avenue for future work.
    \item The brightness temperature of this planet also shows large temporal variations. Other than the case of atmospheres with a sub-stellar temperature of 2500K assumed here, we also showcase the case for a higher substellar temperature of T = 3000K in the Appendix B, which also reiterates the same results about detectabilities presented in this study. However, we do not model atmospheres for lower substellar temperatures in this study, but anticipate lower detectabilities with the same amount of observation time due to reduced emission. Comparing the results from grid based selection is however not trivial because the same oxygen fugacity value results in different types of atmospheres for different brightness temperatures, as can be seen by comparing Figure \ref{abundances} and Figure B2.
\end{itemize}
\subsection{Further Work}
A ground-based observational campaign would be the ideal test for the models developed in this work. Unfortunately, all observations of 55\,Cnc\,e with CARMENES to date have been performed in transmission and are therefore not amenable for the same kind of analysis. To the best of our knowledge, there are no observations in emission using GIANO and SPIRou. CRIRES+ has been used to observe 55\,Cnc\,e in emission in the K band (specifically the K2148 mode) three times, in the very brief period when it can be observed at airmass $<$ 2 during summer in the Southern horizon. However, the first observation did not have the Adaptive Optics (AO) system switched on. The other two (publicly available) observations have exposure times (60s) that would either saturate many pixels in the detectors or have photon counts in many pixels fall above the non-linearity limit according to S/N per pixel calculation with the mode and exposure time equivalent to the two observations with the publicly available online ETC 2.0 calculator. However, 60s is also the lowest recommended exposure time for the Y,J,H and K modes according to the ESO CRIRES+ user manual. So, coupled with the challenge associated with poor AO performance above airmass of 1.4, it is a non-trivial observing scenario for a bright star like 55\,Cnc\,A which makes observation difficult with CRIRES+. Hence, we have decided to not proceed with an analysis of any observational data for the purpose of this study but do plan for, and strongly suggest more observations of this system in emission using ground-based high-resolution spectrographs in the future.
\\
\\
This work on detectability of mineral atmospheres on 55\,Cnc\,e is special because 55\,Cnc\,A is a very bright star. This results in the largest emission spectroscopic index of 55\,Cnc\,e by far ($\sim$ 6 times the next best candidate) among all known magma ocean candidates \citep{seidler2024impact}. Similar analysis for the other magma ocean candidates is hence currently not feasible and will require the use of next generation telescopes like European-ELT, TMT and GMT, which will have much larger aperture sizes. This will enable comparative analysis between all magma ocean world candidates of interest. As also stated before, volatile rich atmospheres are also a possibility, at least for 55\,Cnc\,e, and characterisation of such atmospheres is only possible with larger telescopes. So, future ELTs will have a huge role to play in general for detecting and characterising a wider sample of magma ocean worlds.

\section{Conclusions}\label{conclusion}
USPs like 55\,Cnc\,e being very close to their host star and being tidally locked are expected to have lost their primary atmospheres and have their dayside surface melted to form a Lava/Magma Ocean Planet (MOP) scenario. The secondary dayside atmosphere is then expected to be formed by vaporisation from the surface melt and form mineral atmospheres. The composition of such mineral atmospheres are expected to be very diverse and strongly dependant on the oxygen fugacity ratio of the bulk mantle (rather than just the diversity of the bulk composition itself). Since oxygen fugacity is also an important indicator of the interior composition of the exoplanet, studying the mineral atmospheres of USPs like 55\,Cnc\,e potentially provides a unique way to understand the exoplanetary interiors as well. 
\\
\\
In this first of its kind high-resolution study, we aimed to check if the diverse mineral atmospheres possible due to the bulk mantle oxygen fugacity composition spanning 12 orders of magnitude could be detectable through high-resolution ground-based spectrographs currently mounted at some of the largest observatories on Earth. These atmospheric compositions were obtained using the coupled surface melt-atmosphere model developed in \citet{seidler2024impact} (Figure \ref{abundances}) and the template emission spectra for each atmospheric composition was generated using GENESIS (Figure \ref{emtemplate}). We assumed a substellar temperature of T = 2500K, which falls within the large possible range of brightness temperature of this exoplanet. We also showcase the case for T = 3000K in Appendix B. We construct our own synthetic night simulator \texttt{Ratri} to simulate two nights of observations (11 hours in total) of the dayside of 55\,Cnc\, using CARMENES (Figure \ref{nights}). 
\begin{itemize}
    \item We find that all modelled atmospheric scenarios are detectable using the two simulated nights (Figure \ref{conf066}). This motivated us to then check if all these detectable scenarios could be differentiated from each other, with the difficulty in differentiation arising because all the model templates are overall very similar to each other (Figure \ref{ccf066}).
    \item Performing a test using an optimistic mode of detrending where we can perfectly remove all the telluric absorption and stellar flux, we find that we can differentiate between the all model scenarios and the \fOtwo of the injected model is always correctly retrieved within the $\leq$1$\sigma$ contours (Figure \ref{optsel066}) which means that precise differentiation of all model scenarios is possible.
    \item The default blind SVD/PCA based detrending procedure using the HRCCS analysis pipeline \texttt{Upamana} also closely reproduces the detectabilities expected from the perfect detrending scenario across a wide range of the number of singular vectors used for detrending. This showcases both the efficacy of this detrending framework and the fact that the diverse mineral atmospheres expected on 55\,Cnc\,e are differentiable using 11 hours of dayside observations using CARMENES in particular and any other similar current generation spectrographs in general. This strongly motivates the need for future observations of this exoplanet using ground-based spectrographs to assist in decoding the puzzling nature of 55\,Cnc\,e already apparent from space-based observations.
\end{itemize}
\section*{Acknowledgements}

SD is supported by a Chancellors' International Scholarship from the University of Warwick. PAS and FLS thank the Swiss National Science Foundation (SNSF) through an Eccellenza Professorship (203668) and the Swiss State Secretariat for Education, Research and
Innovation (SERI) under contract No. MB22.00033, a SERI-funded ERC Starting Grant “2ATMO” to PAS. Parts of this work have been carried out within the
framework of the National Centre of Competence in Research (NCCR) PlanetS
supported by the SNSF under grant 51NF40\_205606. We thank the anonymous referee for their comments and suggestions, which helped improve the quality of this manuscript. For the purpose of open access, the author has applied a Creative Commons Attribution (CC BY) licence to any Author Accepted Manuscript version arising from this submission.

%%%%%%%%%%%%%%%%%%%%%%%%%%%%%%%%%%%%%%%%%%%%%%%%%%
\section*{Data Availability}

The pipelines \texttt{Ratri} and \texttt{Upamana}, and the synthetic nights used in this work can be shared upon reasonable request. The pipelines are under active development are are planned to be made open-source in the future.

% The inclusion of a Data Availability Statement is a requirement for articles published in MNRAS. Data Availability Statements provide a standardised format for readers to understand the availability of data underlying the research results described in the article. The statement may refer to original data generated in the course of the study or to third-party data analysed in the article. The statement should describe and provide means of access, where possible, by linking to the data or providing the required accession numbers for the relevant databases or DOIs.



\bibliographystyle{mnras}
\bibliography{bibliography} % if your bibtex file is called example.bib

%\begin{thebibliography}{99}
%\end{thebibliography}

%%%%%%%%%%%%%%%%%%%%%%%%%%%%%%%%%%%%%%%%%%%%%%%%%%

%%%%%%%%%%%%%%%%% APPENDICES %%%%%%%%%%%%%%%%%%%%%
\newpage
\appendix
\section{3-parameter retrievals}
\begin{figure*}
    \centering
    \begin{tabular}{cc}
        (a) $k=4$ & (b) $k=6$ \\
        \includegraphics[width = \columnwidth]{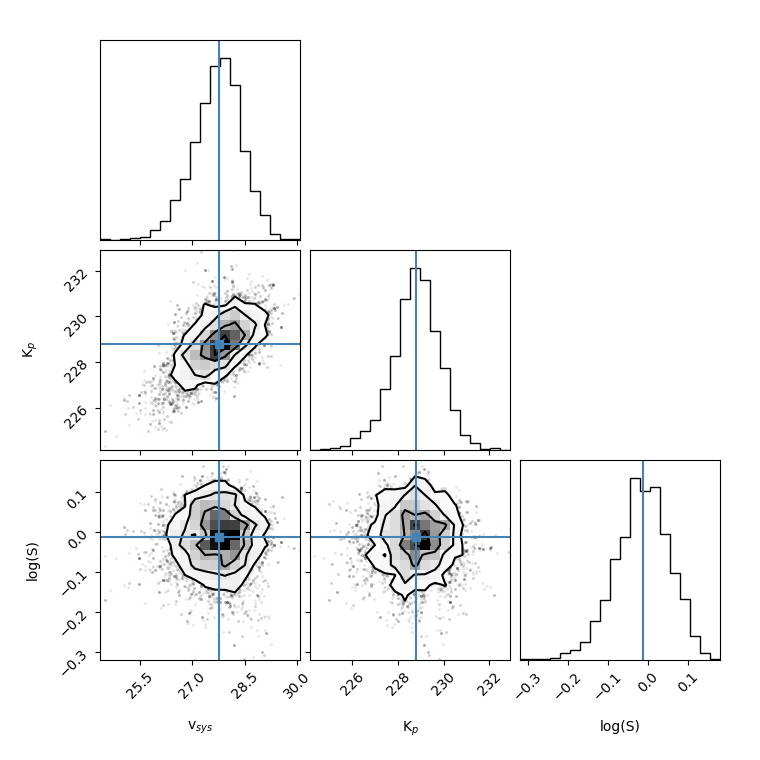}  & \includegraphics[width = \columnwidth]{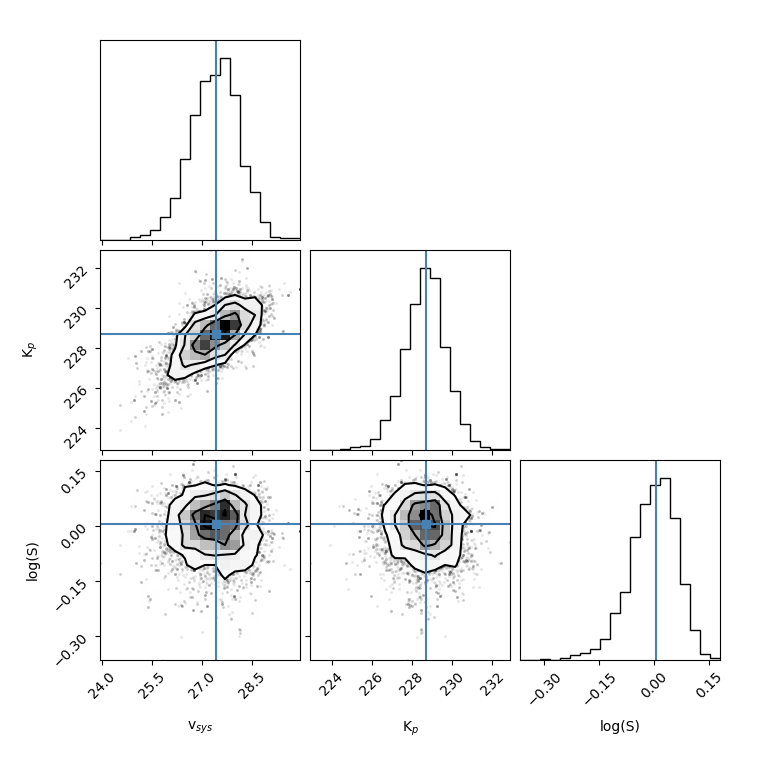} \\
        (c) $k=8$ & (d) $k=10$  \\
        \includegraphics[width = \columnwidth]{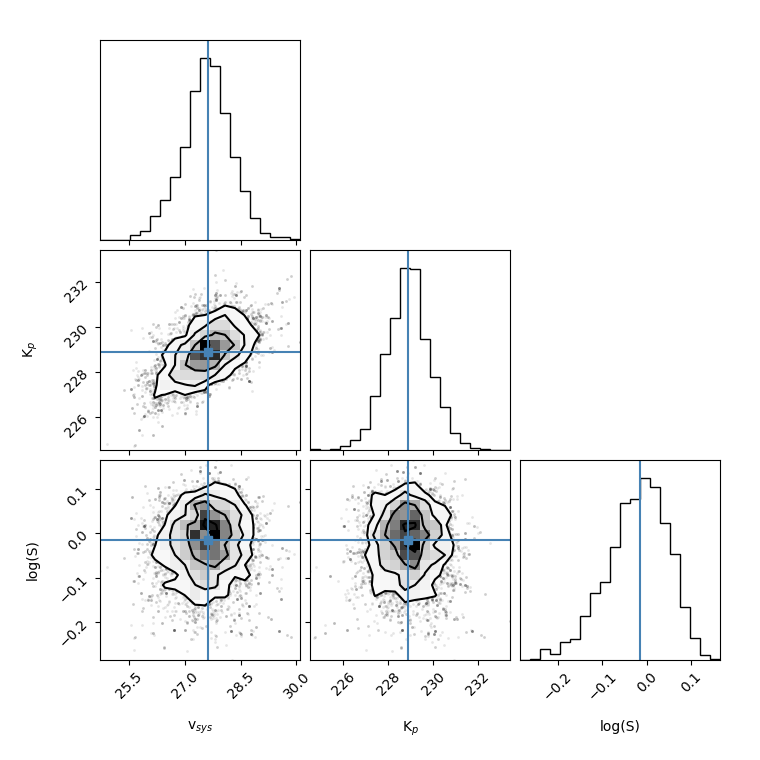} & \includegraphics[width = \columnwidth]{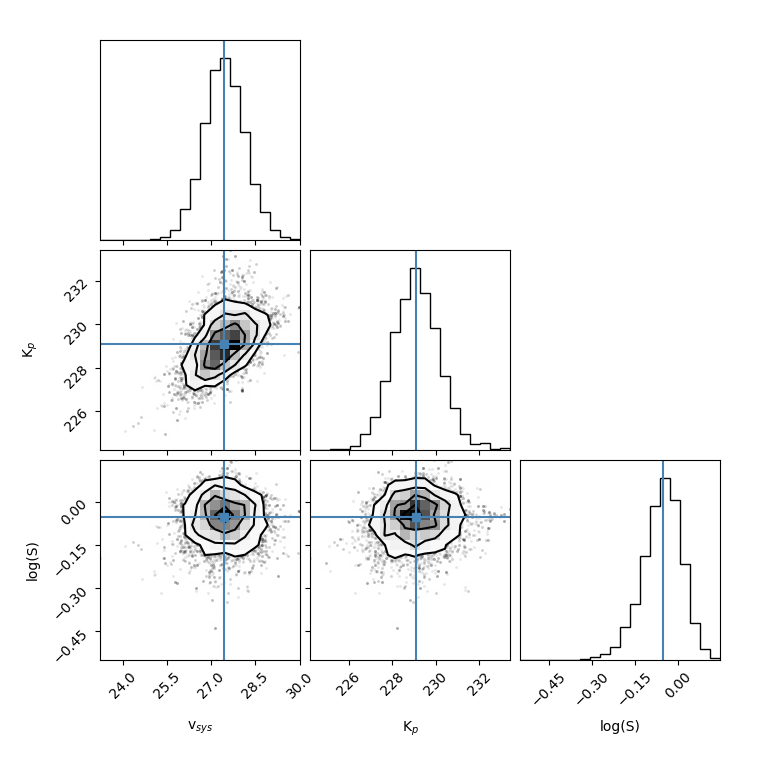} \\
    \end{tabular}
    \caption{MCMC corner plots showcasing successful retrieval within 1000 iterations of the parameters $v_\mathrm{sys}$, $K_\mathrm{P}$ and $\log{S}$ for an injected $\Delta$IW-6 signal in Night 1. Changing the number of SVD components ($k$) used for detrending does not significantly bias the retrieved $\log(S)$, which is always close to 0 i.e. $S$ is close to 1.}
    \label{scalefactor}
\end{figure*}

Since \texttt{Upamana} uses the CCF-to-$\log(L)$ approach similar to \citet{brogi2019retrieving}, it can be coupled to a Bayesian parameter estimation algorithm. For this work, we use the python package \texttt{emcee} for this purpose to retrieve the $v_\mathrm{sys}$, $K_\mathrm{P}$ and $S$ (as $\log(S)$) of an injected signal corresponding to the $\Delta$IW-6 fugacity case in simulated Night 1. This has the strongest line cores as shown in Figure \ref{emtemplate} and is hence expected to be the easiest to detect. We run 1000 MCMC iterations, with the starting 100 iterations serving as a burn-in and disregarded. We repeat the analysis for four different values of $k$ (4,6,8,10) to showcase the effect of the detrending procedure on the retrieval process. The results are shown in Figure \ref{scalefactor}. It is easy to see that the the retrieved $v_\mathrm{sys}$ and $K_\mathrm{P}$ vales are close to the expected values as shown in Table \ref{tab:system_par}. The retrieved scale factor is always close to 1 i.e. 1 lies within 1$\sigma$ of the resulting posterior distribution of $S$. This shows that the assumption of $S=1$ we made in Equation 18 holds for all the values of $k$ considered in this study as part of the grid based model selection process in Section \ref{svddetrend}. It also shows that exoplanet template models considered in this study are expected to be detectable for a wide range of $k$ without suffering from any degradation due to detrending, which is a consequence of both the fast moving nature of the exoplanet, and the faintness of all model template signals.

\section{The case of T$_\mathrm{irr}$ = 3000K}
\begin{figure}
    \centering
    \includegraphics[width = \columnwidth]{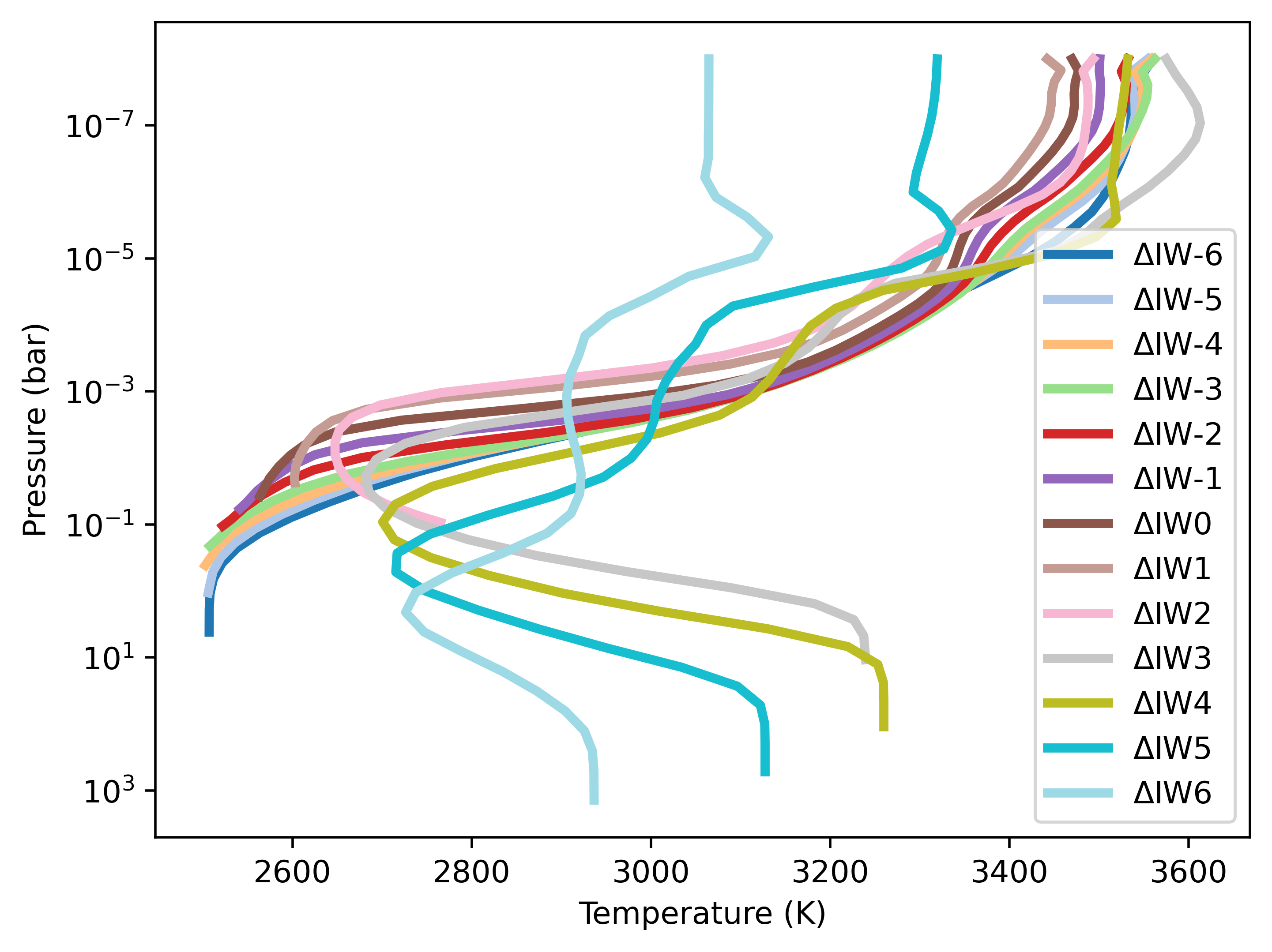}
    \caption{Same as in Figure \ref{ptprof} but for the case of T$_\mathrm{irr}$ $=$ 3000K.}
    \label{ptproff100}
\end{figure}

\begin{figure*}
    \centering
    \includegraphics[width = 2\columnwidth]{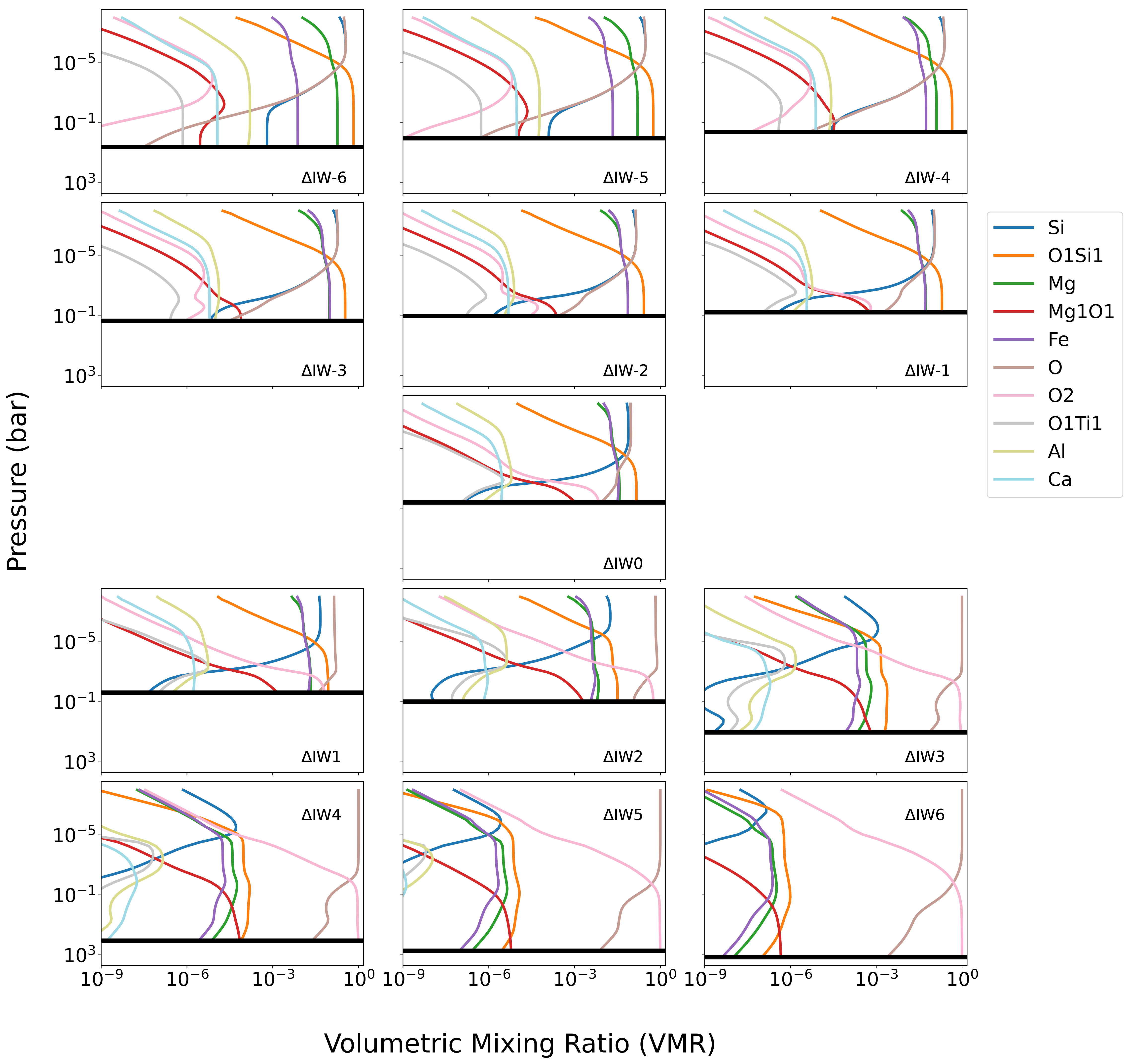}
    \caption{Same as in Figure \ref{abundances} but for the case of T$_\mathrm{irr}$ $=$ 3000K.}
    \label{abundancesf100}
\end{figure*}

\begin{figure*}
    \centering
    \includegraphics[width = 2\columnwidth]{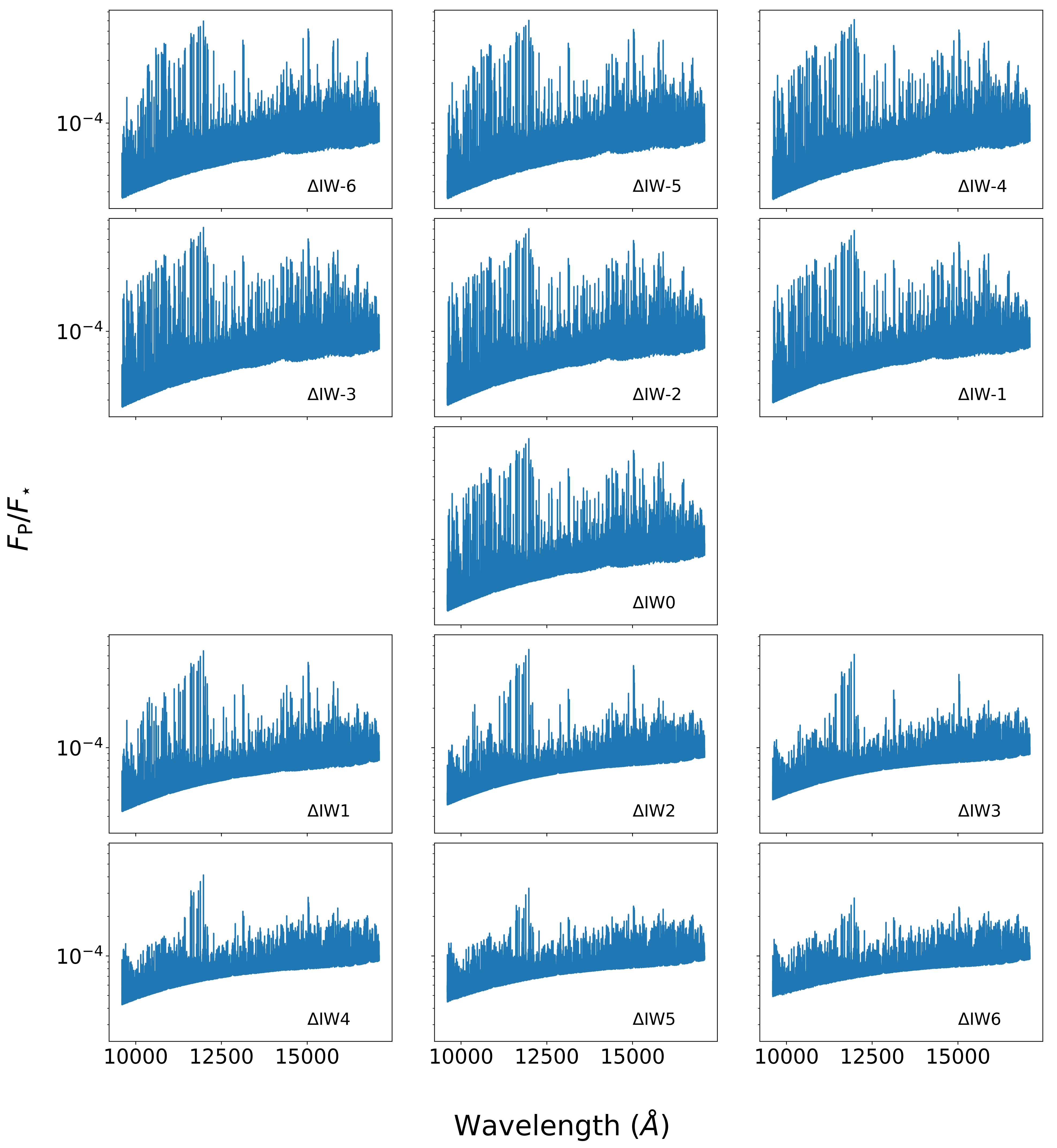}
    \caption{Same as in Figure \ref{emtemplate} but for the case of T$_\mathrm{irr}$ $=$ 3000K.}
    \label{emtemplatef100}
\end{figure*}

\begin{figure}
    \centering
    \includegraphics[width = \columnwidth]{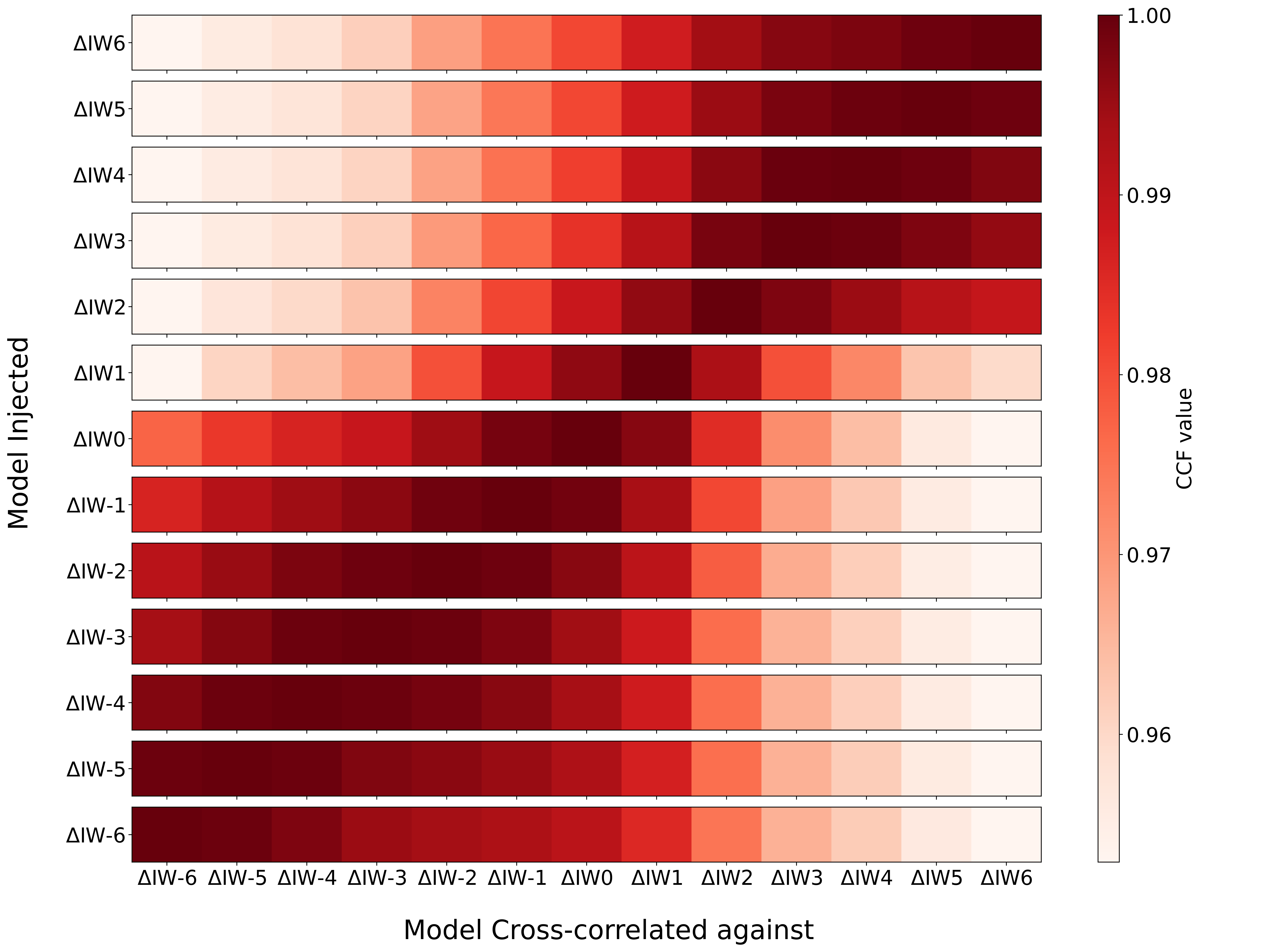}
    \caption{Same as in Figure \ref{ccf066} but for the case of T$_\mathrm{irr}$ $=$ 3000K.}
    \label{ccf100}
\end{figure}

\begin{figure*}
    \centering
    \includegraphics[width = 2\columnwidth]{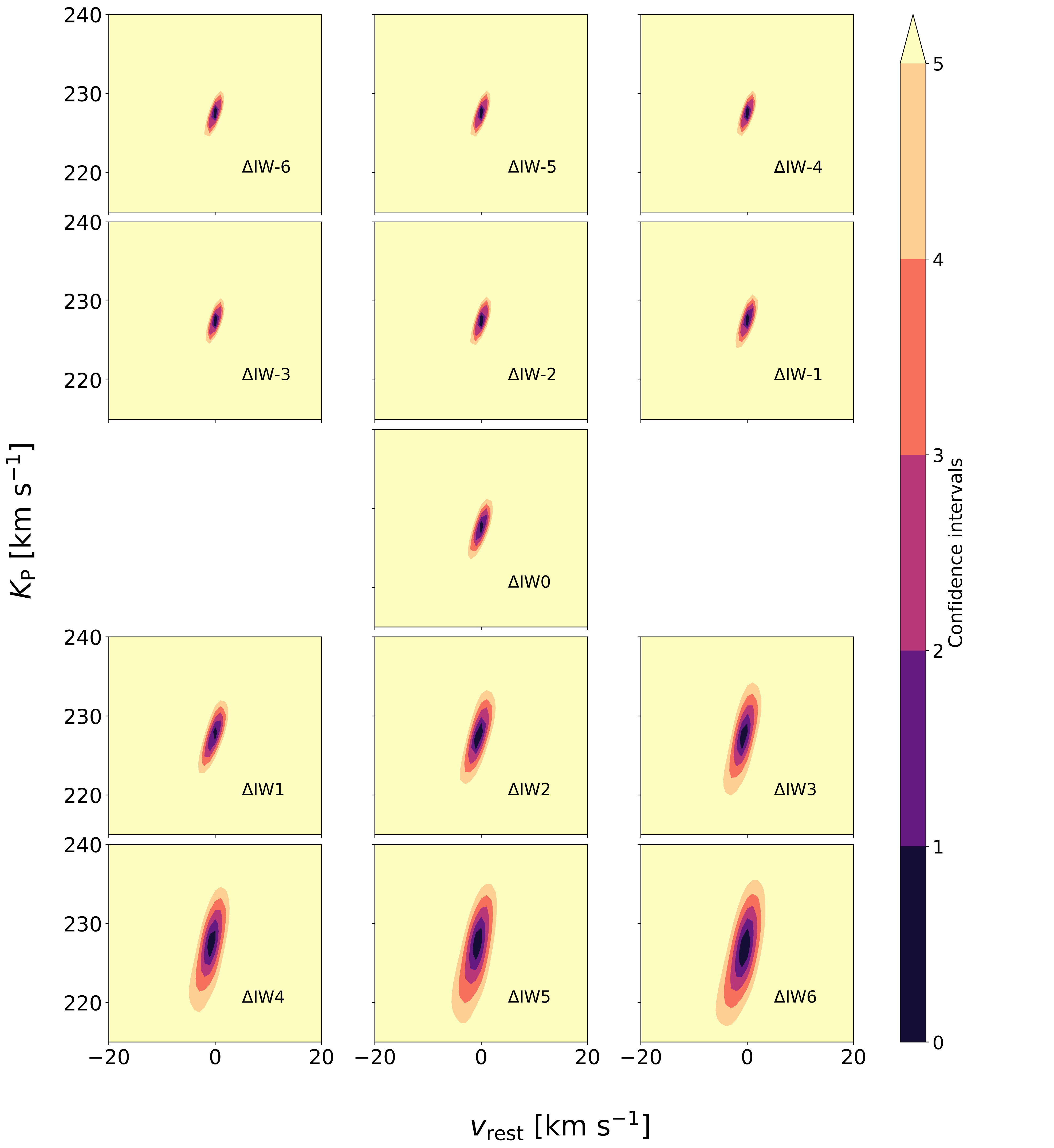}
    \caption{Same as in Figure \ref{conf066} but for the case of T$_\mathrm{irr}$ $=$ 3000K.}
    \label{conf100}
\end{figure*}

\begin{figure}
    \centering
    \includegraphics[width = \columnwidth]{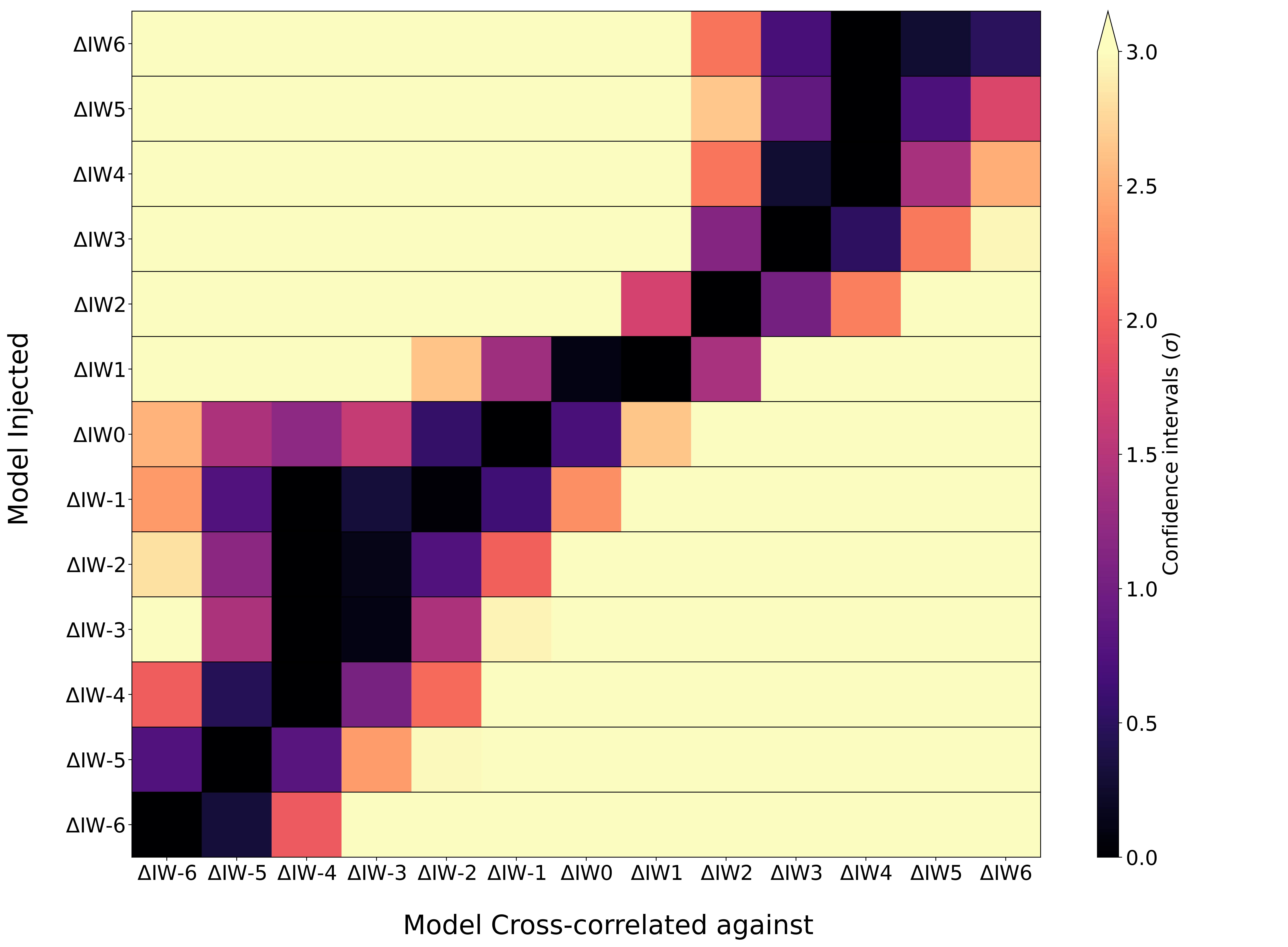}
    \caption{Same as in Figure \ref{optsel066} but for the case of T$_\mathrm{irr}$ $=$ 3000K.}
    \label{optsel100}
\end{figure}

\begin{figure*}
    \centering
    \begin{tabular}{cc}
        (a) $k=4$ for each night & (b) $k=6$ for each night \\
        \includegraphics[width = \columnwidth]{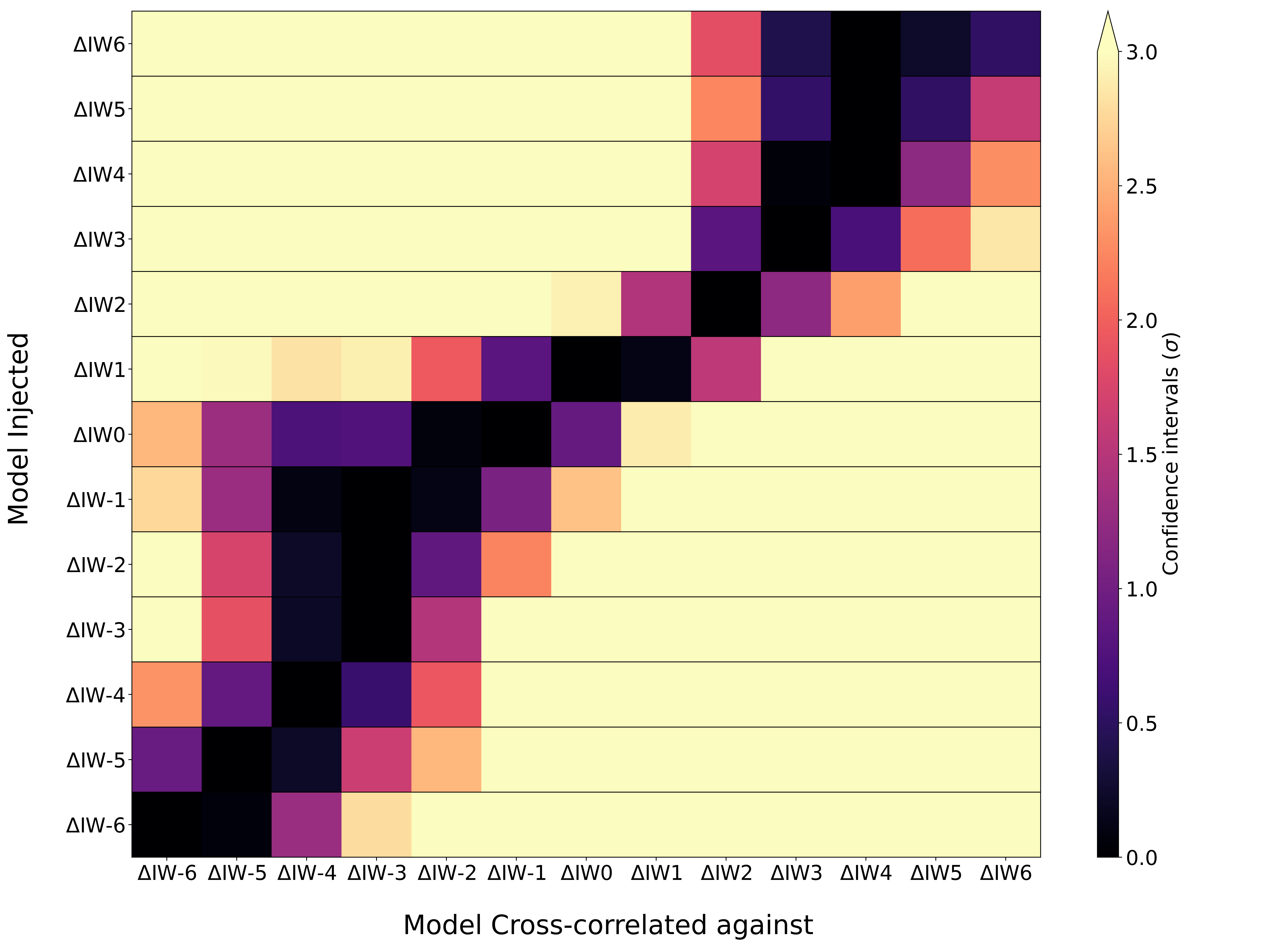}  & \includegraphics[width = \columnwidth]{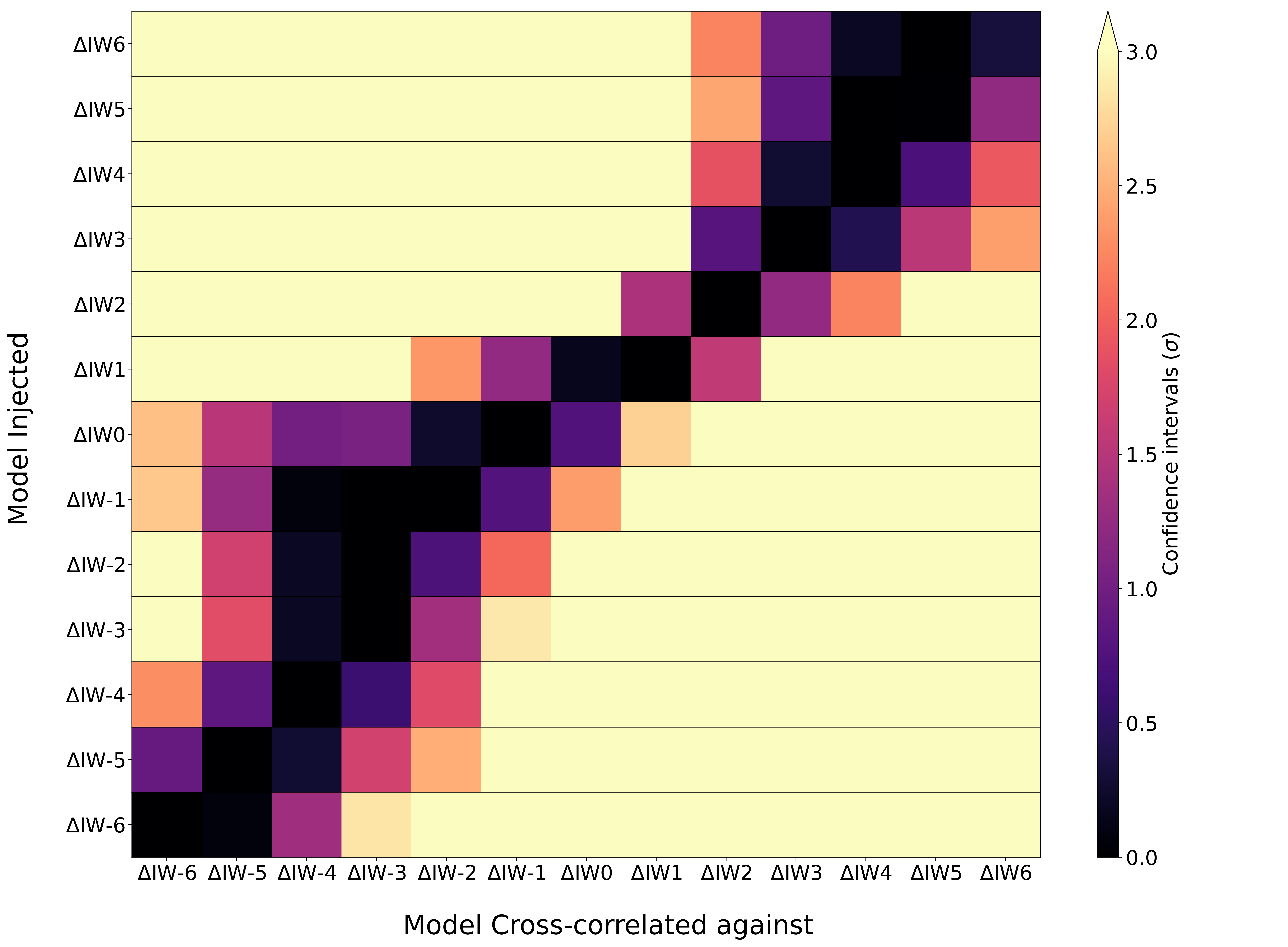} \\
        (c) $k=8$ for each night & (d) $k=10$ for each night \\
        \includegraphics[width = \columnwidth]{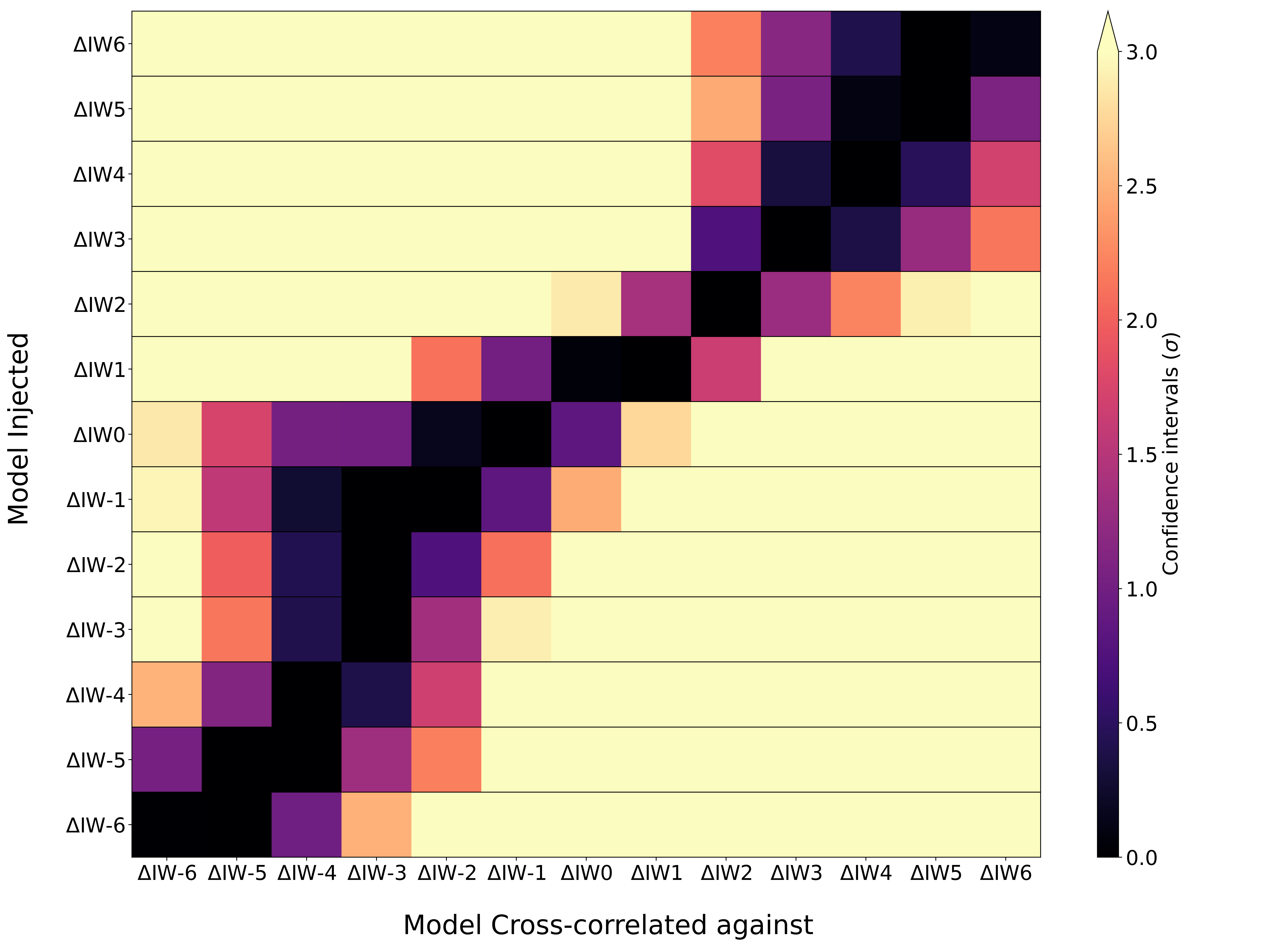} & \includegraphics[width = \columnwidth]{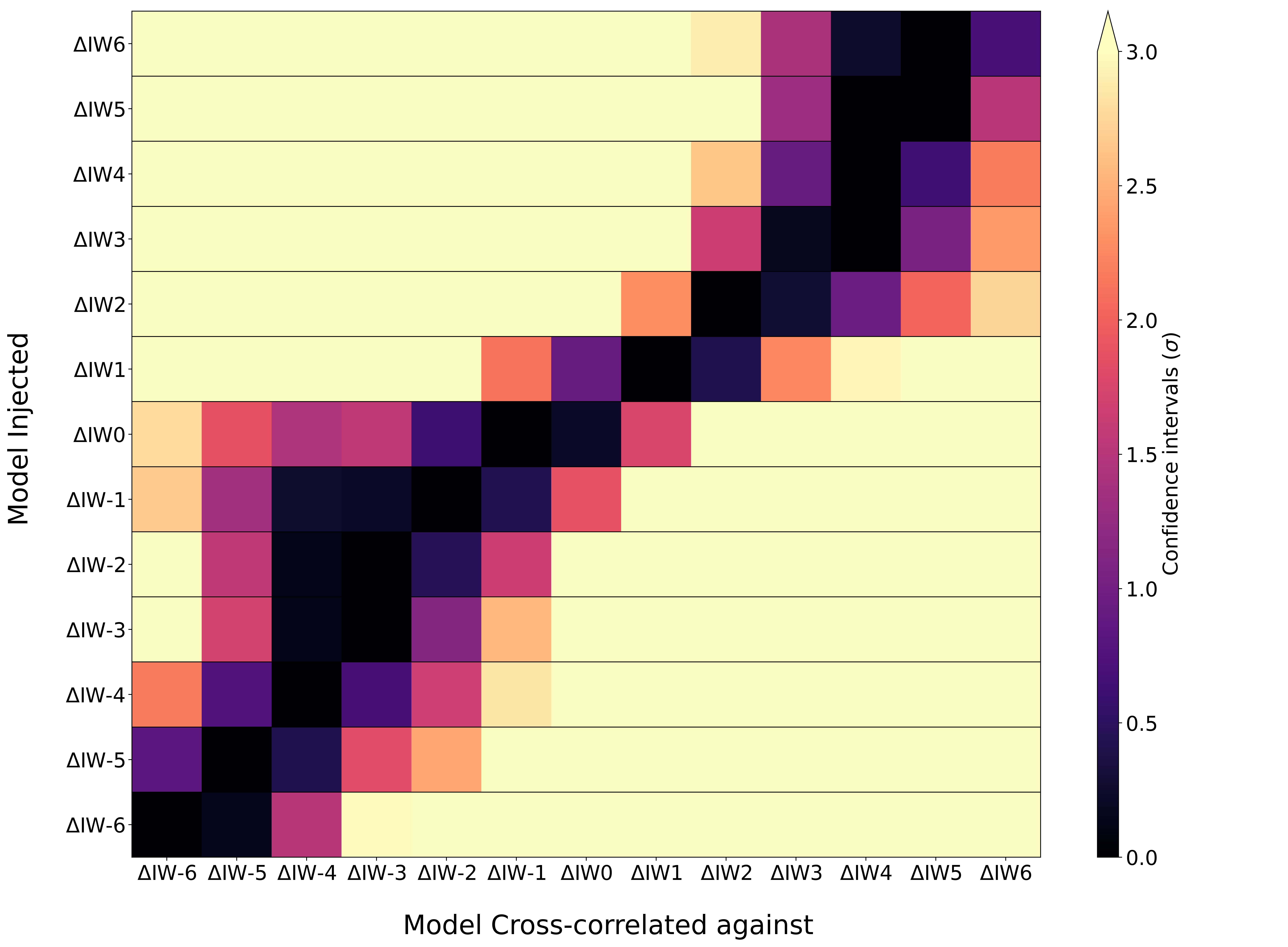} \\
    \end{tabular}
    \caption{Same as in Figure \ref{svdsel066} but for the case of T$_\mathrm{irr}$ $=$ 3000K.}
    \label{svdsel100}
\end{figure*}

The calculation of the substellar/irradiation temperature of the surface magma melt on the exoplanet's dayside is found from the equation \citep{seidler2024impact}:
\begin{equation}
    \mathrm{T_{irr}} = (f(1-A_{b}))^{1/4}\bigg{(}\frac{R_{\star}}{a}\bigg{)}^{1/2}\mathrm{T_{eff}},
\end{equation}
where $A_{b}$ is the Bond albedo, which is a measure of the reflected radiation from the exoplanet. The rest of the parameters are already defined in the main text where we used a value of $f=2/3$. 55\,Cnc\,e showcases a wide range of brightness temperatures and hence we also perform the same HRCCS analysis as in the main text by using a higher value of $f=1$ to get T$_\mathrm{irr}$ of the surface magma melt as 3000K. The new P-T profiles, atmospheric chemical abundances, and emission spectral templates are shown in Figures \ref{ptproff100}, \ref{abundancesf100}, and \ref{emtemplatef100} respectively. All spectral templates show an overall higher value of $F_\mathrm{P}/F_{\star}$ compared to the case of $f=2/3$ due to a greater amount of radiation from a hotter exoplanet. The surface pressures are also higher due to a greater amount of outgassing of all gases. The similarity plot for all these model templates is shown in Figure \ref{ccf100}. As expected of models with stronger line strengths, all these models are also more tightly constrained in Figure \ref{conf100}, compared to the $f=2/3$ case. The perfect detrending based model selection plot (Figure \ref{optsel100}), and all SVD/PCA detrending based model selection plots (Figure \ref{svdsel100}) showcase much tighter constraints around the auto-correlation diagonal compared to the $f=2/3$ case. Hence, these newer models are also detectable and differentiable with 11 hours of dayside observations of 55\,Cnc\,e using CARMENES. 

%%%%%%%%%%%%%%%%%%%%%%%%%%%%%%%%%%%%%%%%%%%%%%%%%%

% Don't change these lines
\bsp	% typesetting comment
\label{lastpage}
\end{document}